\documentclass{sig-alternate}
\usepackage{mathptmx} 

\usepackage{fancyhdr}
\usepackage[normalem]{ulem}
\usepackage[hyphens]{url}
\usepackage[sort,nocompress]{cite}
\usepackage[final]{microtype}
\usepackage[keeplastbox]{flushend}

\usepackage{amsmath,amssymb}
\usepackage{graphicx}
\usepackage{textcomp}
\usepackage{xcolor}
\usepackage{makecell}
\usepackage{multicol}
\usepackage{multirow}

\usepackage[para,online,flushleft]{threeparttable}
\usepackage{longtable}

\usepackage{placeins}
\usepackage{booktabs}

\usepackage[utf8]{inputenc}
\usepackage[ruled,vlined]{algorithm2e}
\usepackage[noend]{algpseudocode}
\usepackage{comment}
\usepackage{breqn}
\usepackage{float}
\usepackage[colorinlistoftodos]{todonotes}
\usepackage{tikz}
\usepackage{epsfig}
\usepackage[center]{caption}
\usepackage{color,soul}
\soulregister\cite7
\soulregister\ref7
\usetikzlibrary{arrows,positioning,shapes.geometric}
\usepackage{tikz}

\def\BibTeX{{\rm B\kern-.05em{\sc i\kern-.025em b}\kern-.08em
    T\kern-.1667em\lower.7ex\hbox{E}\kern-.125emX}}

\fancypagestyle{firstpage}{
  \fancyhf{}
  
  \fancyfoot[C]{\thepage}
}

\title{An Algorithm-Hardware Co-design Framework to Overcome Imperfections of Mixed-signal DNN Accelerators\vspace{-40pt}}
\author{{Payman Behnam, Uday Kamal, Saibal Mukhopadhyay}\\
{Georgia Institute of Technology, Atlanta, USA}\\
{\{payman.behnam, uday.kamal\}@gatech.edu, saibal.mukhopadhyay@ece.gatech.edu}\\}

 \pdfoutput=1
\begin{document}
\maketitle
\thispagestyle{firstpage}
\pagestyle{plain}
\begin{abstract} 
\vspace{-3 ex}
In recent years, processing in memory (PIM) based mixed-signal designs have been proposed as energy- and area-efficient solutions with ultra high throughput to accelerate DNN computations. However, PIM designs are sensitive to imperfections such as noise, weight and conductance variations that substantially degrade the DNN accuracy. To address this issue, we propose a novel algorithm-hardware co-design framework hereafter referred to as HybridAC that simultaneously avoids accuracy degradation due to imperfections, improves area utilization, and reduces data movement and energy dissipation. We derive a data-movement-aware weight selection method that does not require retraining to preserve its original performance. It computes a fraction of the results with a small number of variation-sensitive weights using a robust digital accelerator, while the main computation is performed in analog PIM units. This is the first work that not only provides a variation-robust architecture, but also improves the area, power, and energy of the existing designs considerably. HybridAC is adapted to leverage the preceding weight selection method by reducing ADC precision, peripheral circuitry, and hybrid quantization to optimize the design. 
Our comprehensive experiments show that, even in the presence of variation as high as 50\%, HybridAC can reduce the accuracy degradation from 60 - 90\% (without protection) to 1 - 2\% for different DNNs across diverse datasets. 
In addition to providing more robust computation, compared to the ISAAC (SRE), HybridAC improves the execution time, energy, area, power, area-efficiency, and power-efficiency by 26\%(14\%), 52\%(40\%), 28\%(28\%), 57\%(45\%), 43\%(5$\times$), and 81\%(3.9$\times$), respectively.

\end{abstract}
\section{introduction}
\label{sec:intro}
\begin{figure*}[ht!]
	\vspace{-3ex}
	\begin{center}
		\includegraphics[width =1.8\columnwidth]{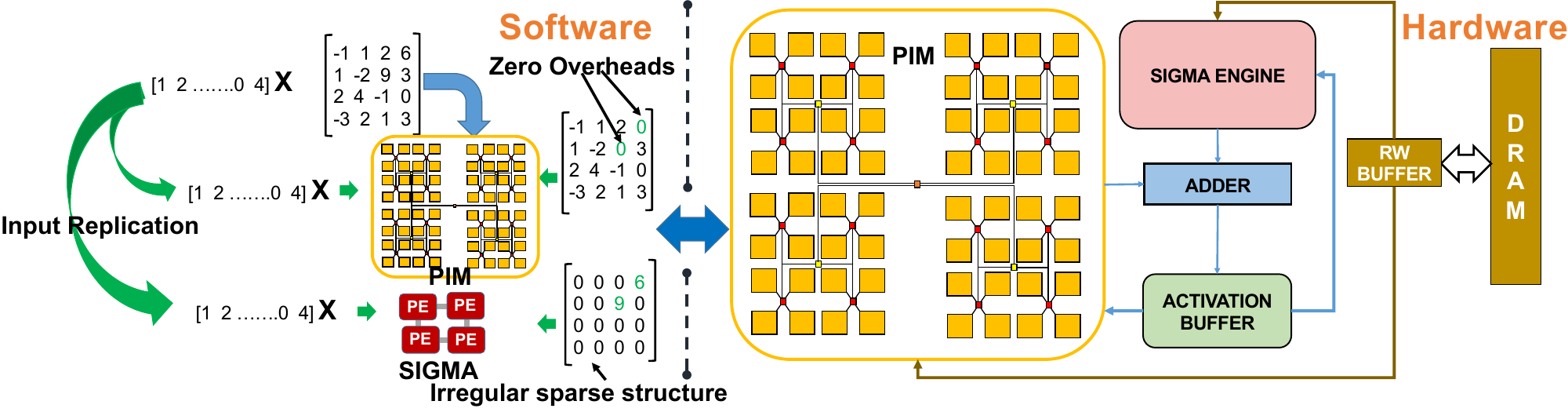}
	\vspace{-2ex}
	\caption{Illustration of the individual weight selection method (IWS) and its corresponding architecture~\cite{Dash2020, dash2021robust}.
		\label{figure:naive.pdf}}
	\end{center}
	\vspace{-5ex}
\end{figure*}

In the recent years, processing in memory (PIM) and near memory (PNM) have been considered as
 fast, energy- and area-efficient solutions to build Deep Neural Network (DNN) accelerators. Among them, ReRAM-based mixed-signal DNN accelerators have
 absorbed a great deal of interest due to their high throughput and energy-efficiency~\cite{shafiee2016isaac, PRIME, Pipelayer, Chen2018, Xue2020, Lu2020, yuan2021tinyadc}. Yet, they are still in the early stages of development since there are several fundamental challenges that need to be addressed properly for their practical applications. 
 
 Firstly, these accelerators are very sensitive to imperfections such as stochastic process variations and  programming errors due to their operations in the continuous analog regimes that degrade their accuracy significantly~\cite{li2017understanding, Chen2018, Chen2018Neuro, rashidi2018improving, Xue2020, Lu2020, Feinberg2018}. Secondly, since ReRAM cells are non-ideal devices, even if the cells store the exact weight values, 
 the current on each bit line can deviate from the actual current that was supposed to flow especially in the architectures like ISAAC that exploits bias and offset subtraction for mapping the weights~\cite{shafiee2016isaac}. To have a higher throughput, more wordlines needs to be activated concurrently. By activating more wordlines, the amount of deviation in accumulated current, which is the input of analog to digital converter (ADC) becomes larger. This leads to an overlap of neighboring states in ADC's output and error in the computation~\cite{Chen2018, RSE2019}.  
  In practice, we cannot activate multiple wordlines (i.e, greater than 16) at the same time, which results in significant throughput degradation~\cite{yuan2021forms, Chen2018, RSE2019}.

Thirdly, the employed ADCs in the mixed-signal accelerators consume significant area and power consumption (e.g., between 30\% to 50\% of the area, and between 50\% to 70\% of power consumption)~\cite{shafiee2016isaac, PRIME, Wu2020,imani2019floatpim}. Some designs leverage smaller ADCs to perform the computations in lower-bit precision to save area and energy consumption at the cost of the undesired accuracy drop~\cite{Lu2020}. These properties make these mixed-signal DNN accelerators an unsuitable choice for many existing applications.

Recently, several works have proposed solutions to make these accelerators more reliable~\cite{Dash2020, dash2021robust, Long2019, Feinberg2018, Chang2017,Kazemi2020, liu2019fault,jain2019cxdnn,yang2021multi,ali2020erdnn, li2017understanding}. The closest related work was done by Dash et al.~\cite{Dash2020, dash2021robust}, where they present an algorithmic solution to increase reliability as shown in \textbf{Figure~\ref{figure:naive.pdf}}. They present an approach that selects a fraction of important weights using a hessian-based solution and put them in digital units without considering its impact on the hardware cost. We call this approach IWS.
Relying on this approach leads to a weight selection with a huge irregular sparsity. Supporting this type of sparsity needs not only considerable hardware overhead but also significant data movements that results in a design with high area, power, and low energy efficiency~\cite{behnam2018r, behnam2019stfl, behnam2020stfl}. For instance, IWS suggests using \textit{expensive SIGMA~\cite{qin2020sigma}} design as a centralized digital accelerator to perform sparse computation assigned to the digital cores. However, SIGMA exploits a configurable systolic array-based architecture with remarkable interconnection overheads. Most importantly, \textit{the same input data/activation needs to be replicated in both analog and digital units} to support correct computation of both sensitive and non-sensitive weights even if a very small portion of the data is located in digital cores, which leads to huge data movement and hence energy consumption (\textbf{Figure~\ref{figure:naive.pdf}}). 
IWS also recommends writing weights of the next layers in the same ReRAM crossbars after the finishing the current layer. That means for each layer, even if the analog unit is finished earlier, it needs to wait for a long time until the slower digital unit has finished its computation. In addition, writing into the same ReRAMs per layer is time/energy consuming and has endurance issue~\cite{grossi2019resistive, kondratyuk2021automated, chen2020reram}. Moreover,  weights transferred from analog to digital units leave zeros in their original positions that leads to additional memory overhead. Those zeros suffer from variations as well.

In this work, we propose HybridAC, an algorithm-hardware
co-design framework that not only addresses the aforementioned problems of the baselines~\cite{Dash2020, dash2021robust}, but also provides more opportunities to optimize area, power and energy consumption of the PIM-based mixed-signal accelerators.

\begin{figure*}[ht!]
	\vspace{-3ex}
	\label{ChanelWiseAlg}
	\begin{center}
		\includegraphics[width =1.8\columnwidth]{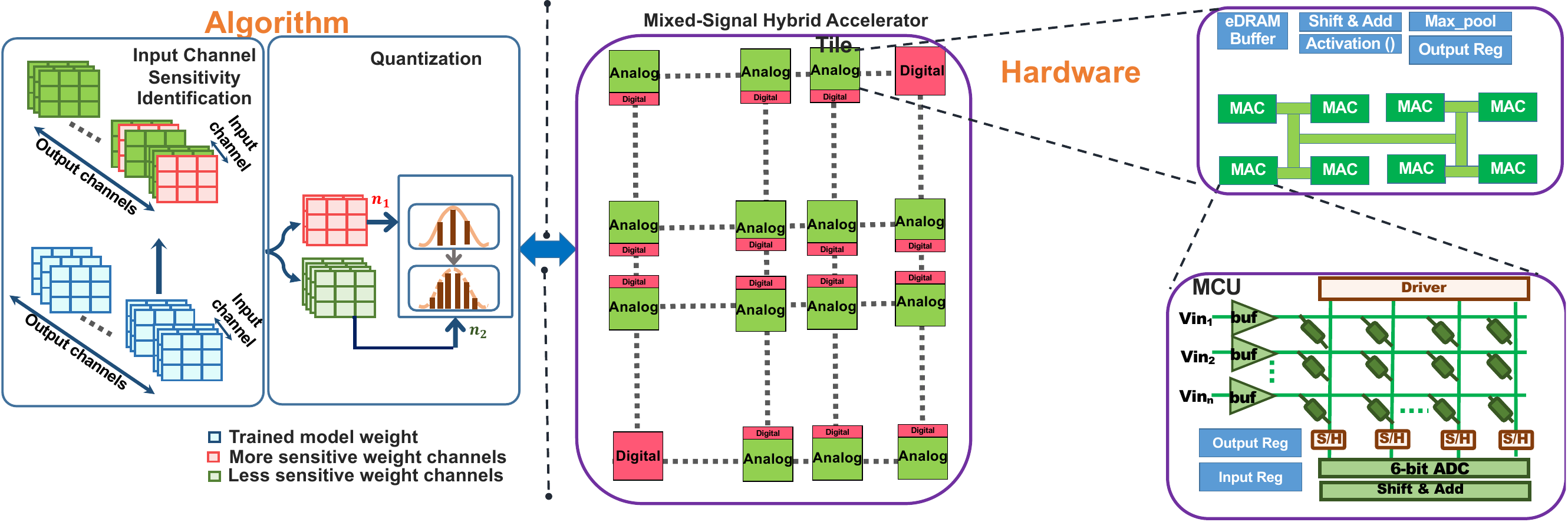}
	\vspace{-3ex}
	\caption{Illustration of the proposed input channel-wise weight selection.
		\label{figure:ChanelWiseAlgWHard}}
	\end{center}
	\vspace{-5ex}
\end{figure*}

HybridAC proposes a \textbf{novel data-movement-aware weight selection} method that captures a fraction of critical weight channels along with their corresponding input channels (e.g., the weights channels that are sensitive to variation and significantly contribute to the final accuracy) per layer and performs their associated computations in the proposed digital accelerator
(\textbf{Figure~\ref{figure:ChanelWiseAlgWHard}}).

The channels are mapped to the rows of crossbars in the mixed-signal designs. Since important weight channels are entirely selected to be placed in digital units,  their corresponding rows in analog crossbar units are removed uniformly. This feature evenly reduces the aggregated partial sums over the bit lines. Hence, we are able to \textbf{employ low precision ADCs} that significantly reduce area and power consumption with a negligible accuracy loss~\cite{shafiee2016isaac,imani2019floatpim}. Moreover, channel-wise sensitive weight selection enables us to quantize the weights in the analog part with lower precision compared to the digital cores \textbf{without any need to retraining/post-training} while maintaining minimal loss in accuracy.
  
In summary, HybridAC brings the following contributions:

\begin{itemize}
\item Unlike IWS~\cite{Dash2020, dash2021robust}, the proposed channel-wise weight selection method \textbf{avoids replicating the input activation} and significantly reduces data movement and energy consumption.

\item In contrast with the 
 previous works~\cite{Long2019, jain2019cxdnn}, the proposed method \textbf{doesn't need any retraining}. This is important as training is a costly procedure and it may not capture all kinds of variations that may occur during run-time operation.
 
\item The proposed solution \textbf{avoids adding extra zero weights in place of transferred weights}, as the whole weight channel along with its corresponding input activation are transferred to the digital units. This reduces hardware overheads compared to the IWS solutions that select important weights without considering the hardware cost~\cite{Dash2020, dash2021robust}.

\item 
For the first time, we propose a \textbf{novel hybrid quantization method for different input channels} per layer by considering whether the weights are mapped to digital or analog cores, without requiring any fine-tuning.

\item Due to transferring important weight channels to digital cores uniformly, HybridAC enables us to utilize \textbf{smaller ADC and peripheral circuitry}. In conjunction with hybrid quantization, HybridAC is able to \textbf{further reduce  the area, power, and energy consumption of analog part significantly}, where a fraction of the saved area can be utilized to add a \textbf{proposed digital accelerator} that is robust against variation and protects the sensitive weights.

\item Since HybridAC provides a design robust to the conductance variation, we are able to \textbf{activate more wordlines of the analog crossbar at the same time} without being worried about the accuracy degradation. 
\end{itemize}

\section{Proposed Mechanism}
\vspace{-2ex}
\subsection{Input Channel-wise weight Selection}
To address the above mentioned problems, we propose HybridAC --an end-to-end hardware-software co-design framework-- to overcome the inherent imperfection of the mixed-signal DNN accelerators that includes several innovations in both software and hardware perspectives (Figure ~\ref{figure:ChanelWiseAlgWHard}).

Generally, the input to a CNN model is a 4D tensor
($B\times H \times W \times C$), where $B, H, W,$ and $C$ represent the batch size, height, width, and the number of channels, respectively. Similarly, weight parameters consist of $K$ number of 3D \emph{kernel tensors} of dimension ($R\times R \times C$), where $K$ is the number of output channels. $R$ and $C$ denotes height (and width), and depth of the kernel which is equal to the number of channels of the input activations. We propose an algorithm that extracts weights channel-wise that are sensitive to imperfections. We then map these weights in the digital units and the rest in analog units.
Algorithm~\ref{algorithm: WeightSelectionAlgo} and Figure~\ref{figure:ChanelWiseAlgWHard} show the proposed solution. We take already trained network parameters as input, and generate a binary mask that assigns either 0 or 1 values to all the input channels of every kernels in the network, where 1 represents that channel needs to be protected. That means we need to map it to the digital accelerator to achieve the desired test accuracy. In order to compute this mask, we need to take into account the sensitivity of each parameter with respect to the variation. The sensitivity can be calculated using either the gradient~\cite{grad1, grad2} or the Hessian~\cite{dash2021robust, Dash2020} of the parameters. Current research shows Hessian-based solutions reach superior accuracy in terms of estimating the robustness or sensitivity of the parameters~\cite{dash2021robust}. Following that, we calculate the weight sensitivities using the equation~\ref{eqn:sensi} as outlined in~ \cite{dash2021robust}:
\begin{equation}
\label{eqn:sensi}
s = \sum_{i=1}^{n}(|\lambda_i|q_i^2)\cdot w^2 
\end{equation}
Where $w$ is the parameter matrix and $\lambda_i$, $q_i$ are the eigenvalue and the corresponding eigenvector of the Hessian of the parameter matrix with respect to the training objective, and $n$ is the number of top-eigenpairs that are considered. After calculating the summation, we estimate $s$  by taking the Hamdard product of the obtained \textit{sum} and the weight matrix, \textit{w}. 
Next, in order to reduce the redundant data-movement and avoid input data repetition, instead of considering the sensitivity values of each parameter independently as suggested in previous works~\cite{dash2021robust, Dash2020}, we propose input channel-wise aggregated form. 
If P and Q are number of input channels found to be more and less sensitive to variation, respectively, where Q = C – P , then we can say the number of parameters mapped to the digital and analog accelerators for this specific kernel will be $W_d \in R_{R \times R \times P \times K}$, and $W_a \in R_{R \times R \times Q \times K}$, respectively. The sensitivity of a particular input channel is calculated based on the aggregation of all the parameters sensitivity values along that dimension as equation~\ref{eqn:aggsensi} shows: 
\begin{equation}
\label{eqn:aggsensi}
 s_i= \sum_{K}\biggl(\sum_{R}\biggl(\sum_{R}s\biggr)\biggr)
\end{equation}
Where $s_i$ is the aggregated sensitivity values for the \emph{W} kernel. The proposed algorithm computes $s_i$ value for all channels of all existing layers of the given DNN and sorts them according to their magnitude. Initially, it starts its computation by assuming that all channels are mapped on the analog units. Then, it applies variation to all of the weights and computes the new accuracy, $ACC_{calculated}$. If $ACC_{calculated}$ is less than $ACC_{desired}$, it selects the top channel of the sorted list of channel-wise sensitive values and puts it in the digital unit to be protected from the injected noise. We repeat this procedure iteratively until we reach the desired accuracy. The output of this procedure would distinguish the channels that need to be mapped to the analog and digital units and hence the corresponding channels of input data as shown in the left side of Figure~\ref{figure:ChanelWiseAlgWHard}.\footnote{we tried different methods like maxing, averaging, mean square error, etc. Among them, aggregating has been found to provide better accuracy.}

Compared to the individual weight selection~\cite{Dash2020, dash2021robust}, the proposed solution reduces extra data  movement and avoid irregular and expensive sparsity.   
This method also abstains from inserting extra weights with zero values in the place of transferred weights because the entire weight channel is moved to the digital units. These improvements significantly contribute in reducing energy consumption and hardware overheads.

\begin{algorithm}[h!]
\SetAlgoLined
\KwResult{Analog and Digital mapping of the channels}
\textbf{Input}: Trained weights, $\lambda_i$ eigenvalues and $v_i$ corresponding eigenvectors, $ACC_{desired}$, 

\textbf{Calculate} sensitivity of each parameter using: $s_{w} \equiv\left(\sum_{i=1}^{n}\left|\lambda_{i}\right| q_{i}^{2}\right) \odot w^{2}$\;
$SOL =$ [ ]\;
\For{$i<num\_layers$}{
\For{$j<num\_channels$}{
  $SOL_{i_j}$ = channel wise aggregation of sensitivity values\;
  $SOL.append(SOL_{i_j})$
 }
 }
$SOL.\textbf{sort()}$\;
$Digital_{channel}=$ [ ]\;
$Analog_{channel}= SOL$ \;
\textbf{Apply} $noise$ in $Analog_{channel}$\;
$ACC_{calculated}$ = \textbf{Inference($Digital_{channel}, Analog_{channel}$)}\;
\While{$ACC_{calculated} < ACC_{desired}$}{
$channel = Analog_{channel}.\textbf{pop()}$\;
$Digital_{channel}.\textbf{push}(channel)$\;
\textbf{Apply} $noise$ in $Analog_{channel}$\;
$ACC_{calculated}$ = \textbf{Inference($Digital_{channel}, Analog_{channel}$)}\;
 }
\caption{Input Channel-wise Weight Selection}
\label{algorithm: WeightSelectionAlgo}
\end{algorithm}

\subsection{Hybrid Quantization}
\label{sec:quant}

Quantization is a well-known hardware-friendly solution to reduce computation efforts and increase energy efficiency. After the quantization, the original weights and/or activations are represented with lower precision as fixed-point or integer representation while the accuracy is preserved.  
If we represent the original floating-point tensor by $x_f$, its quantized by $x_q$, the zero-point by $zp_x$, the scaling factor by $s_x$, and the number of bits used for quantization by $n$, we will have~\cite{nzmora2019distiller}:
\begin{equation}
\begin{split}
x_q =round((x_f -min{x_f}) \frac{2^n-1}{max_{x_f}-min_{x_f}})= \\round(x_fs_x- min_{x_f}s_x)=round(x_fs_x - zp_x)
\label{eq:gq}
\end{split}
\end{equation}
As shown, the scaling factor is a parameter that depends on the desired bit-precision, \emph{n}, and the range of tensor values ($max-min$).
In a typical CNN, each output feature map for a particular layer is generated by convolving the input feature maps with a number of filters. 

\begin{equation}\label{eq:conv}
y = \sum_{0}^{M-1}w_{m}*x_{m}
\end{equation}

where $M$ is the number of input channels for the kernel of that layer and $*$ denotes convolution operation between input $x_m$ and $w_m$ as 2D tensors.

The linear quantization approximates the results as follows:
\begin{equation}\label{eq:main_eq}
  \begin{aligned}
    y_{q}=\operatorname{round}(\frac{s_{y}}{s_{x} s_{w}}(\sum(x_{q}+zp_{x})(w_{q}+z p_{w}))) 
  \end{aligned}
\end{equation}

where $y, x, w, $ represent the output activation, input activation, and weight respectively. The other parameters that are $s, q, z_p$ represent the scaling factor, quantized output, and the zero-point quantization offset, respectively. 

In HybridAC, the input channel with important weights are steered to digital cores while the rest of them are fed into analog ones. Since analog and digital cores do not share any weights, they can have a different number of quantization bits and different scaling factors. Although digital and analog cores do not share activations, we consider a similar quantization for all activations. This is because one tensor can be an input to multiple layers in some DNNs (e.g., DenseNet121). The partial results of the analog and the digital cores are only added once at the end for each layer:
\begin{equation}\label{eq:conv_mod}
y = y_d + y_a = \sum_{d=0}^{D-1}w_{d}*x_{d}+\sum_{a=0}^{A-1}w_{a}*x_{a}
\end{equation}

where $A$ and $D$ represent the number of input channels computed inside analog and digital cores and $D + A = M$ is total number of input channels. HybridAC quantizes the weights in analog with $n_1$ bits and weights in digital with $n_2$ bits, where $n_2 > n_1$ as important channels are processed in digital cores. 

In order to add the analog and digital partial results, we need to have them in the same scaling factor. So, before adding them together, partial results are converted back to floating points, as follows.  
\begin{equation}\label{eq:main_eq2}
  \begin{aligned}
    y_{fd}=\frac{s_{y}}{s_{x} s'_{w}}(\sum_{d=0}^{D-1}(x_{qd}+zp_{x})(w'_{qd}+z p'_{w}))
  \end{aligned}
\end{equation}

\begin{equation}\label{eq:main_eq3}
  \begin{aligned}
    y_{fa}=\frac{s_{y}}{s_{x} s''_{w}}(\sum_{a=0}^{A-1}(x_{qa}+zp_{x})(w''_{qa}+z p''_{w}))
  \end{aligned}
\end{equation}
where $s'_w$ and $s''_w$ are digital and analog scaling factors, $w'_{qd}$ and $w''_{qd}$ are digital and analog quantized weights, and $zp'_w$ and $zp''_w$ are zero points in digital and analog domains. 
Instead of rounding $y_{fd}$ and $y_{fa}$ and then summing them up, HybridAC first adds the floating-point values and then applies the rounding operation in order to reduce the quantization error. 
Note that converting from quantized value to floating-point happens in all typical DNNs in the return path before feeding the results into the next layer~\cite{jacob2018quantization, shafiee2016isaac, dash2021robust}. 
Here an extra floating-point addition is required compared to the linear quantization described in equation~\ref{eq:main_eq}. We also consider the area and power overhead of the quantization circuits in the estimated results.

We use FP16 data type for $y_{fd}$ and $y_{fa}$ as prior work shows that FP16 accumulator is enough to preserve the accuracy during inference~\cite{abdelaziz2021rethinking}. Our simulation results show that if the weights in the analog and digital accelerators are 6-bit and 8-bit respectively, there would be negligible accuracy loss compared with the case that both of them are in 8-bit. The main advantage of the quantization is that it will reduce the number of required ReRAM cells theoretically by a factor of $\frac{8}{6} = 1.33\times$ that will lead to less resource consumption and computation efforts. To the best of our knowledge, this is the first work that proposes hybrid quantization for different input channels per layer depending upon where those channels are placed. The proposed method \textit{does not require any post-quantization fine-tuning}, as it causes minimal accuracy degradation. 
\section{Architecture Building Blocks }
\label{sec:BB}

\subsection{Analog Accelerator Units}
\label{sec:BBAnalog}
As the right side of Figure~\ref{figure:ChanelWiseAlgWHard} shows the analog/digital units, called analog/digital tiles, are connected with an on-chip mesh fashion. Each analog tile is composed of several multiply-accumulate units (MACs), an eDRAM buffer to
store input values, shift-and-add units that compute final results, output registers to store computed results,  max-pooling and non-linearity activation units that are connected with a shared bus. MAC units include inverters, multiple crossbar subarrays, ADCs, and shift-and-add units. 
Existing architectures~\cite{shafiee2016isaac, PRIME} either propose to activate many wordlines at the same time and use large ADCs (i.e, 8-bit) to achieve high throughput (which is not feasible in practice due to its undesired side-effects as we mentioned in Section~\ref{sec:intro}), or they propose to use smaller ADCs at the cost of having lower accuracy~\cite{Lu2020}, or even activating few number of wordlines, which leads to significantly reduced throughput~\cite{yuan2021forms, RSE2019}. In HybridAC, we are able to activate many wordlines at the same time while using smaller ADCs with small accuracy degradation.

In the proposed architecture, channels are mapped to the rows of crossbar arrays. Since important input channels are mapped to the digital units, those rows will not exist anymore in the analog units. On the other hand, ADCs are located per bitlines. Although only small portion of the weights are mapped to the digital accelerator, they play crucial roles in the final accuracy. Removing rows with important weights causes the value of accumulated current per column to be less sensitive to the variation. Our experimental studies show that if we use ADCs with a lower precision for the crossbars where none of the rows are removed, there would be considerable accuracy drop. However, HybridAC allows us to use ADCs with lower resolution that enables exponential reduction in the area/power~\cite{saberi2011analysis}.

Since partial sum values over the bitlines are reduced, we are now able to employ smaller sample-and-hold and shift-and-add units. Moreover, when a channel is mapped to digital units, the associated rows are completely removed from analog crossbars. This feature prevents from having redundant zeros in the analog cores that existed in the prior works~\cite{Dash2020, dash2021robust}. When a percentage of the weights are placed in digital accelerators, we need fewer crossbars holding the weights. Most importantly, since noise and variation have less impact on the remaining weights on the analog units, we can activate more wordlines at the same time with less concern on the accuracy degradation.
These features are not feasible in the baselines (i.e., IWS)~\cite{Dash2020, dash2021robust} as the locations of important weights are randomly scattered throughout the crossbar bitlines. 
This randomness averts from proposing any scheme that works for all world lines and bitlines uniformly (e.g., exploiting smaller ADCs and peripheral circuits, activating more wordlines).

\subsection{Digital Accelerator Units}
\label{sec:BBDigital}

\begin{figure}[b!]
	\vspace{-3ex}
	\begin{center}
		\includegraphics[width =0.8\columnwidth]{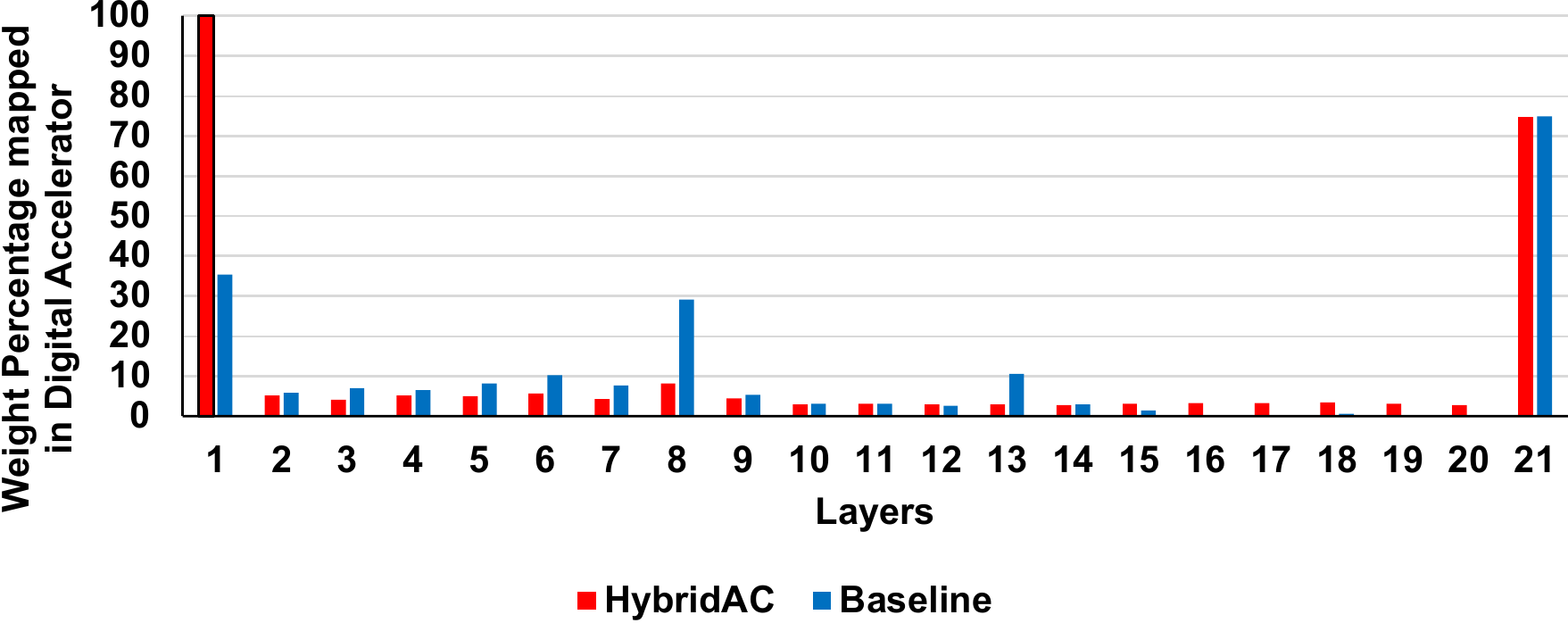}
	\caption{Comparing distribution of the selected important weights in ResNet18-CIFAR10 (HybridAC vs. IWS~\cite{Dash2020})
	\label{figure:distResNet18CIFAR10}}
	\end{center}
	\vspace{-3ex}
\end{figure}

To have a better understanding of how the architecture of a digital unit at the high level looks like, we need to have an understanding of distribution of the selected weights mapped to the digital accelerator among assorted layers. Figure~\ref{figure:distResNet18CIFAR10} illustrates these results for ResNet18 over CIFAR10 data set. We observe the deviation of important weights per layer in HybridAC is less than the IWS method~\cite{Dash2020, dash2021robust}.
\begin{figure}[ht!]
	\vspace{-3ex}
	\begin{center}
		\includegraphics[width =0.8\columnwidth]{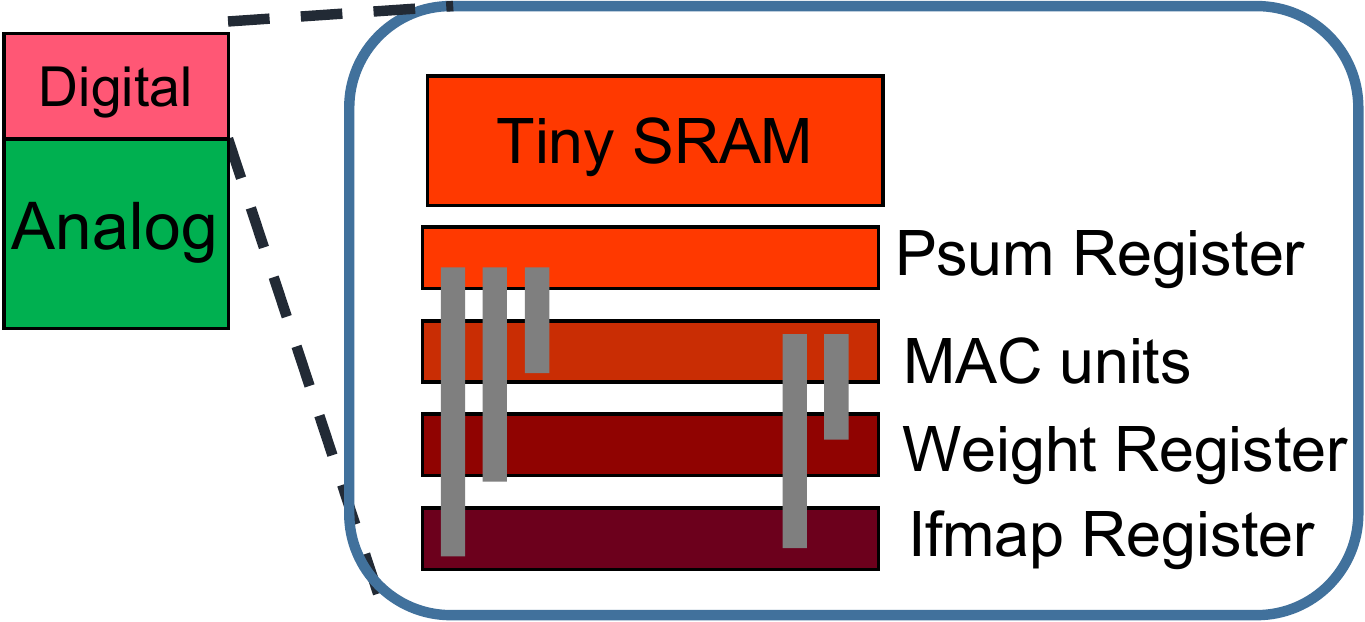}
	\vspace{-2ex}
	\caption{Overall view of a digital core building block
		\label{figure:DigitalUnit}}
	\end{center}
	\vspace{-6ex}
\end{figure}
As seen, for the first and the last layers, both individual and channel-wise wight selection methods map the significant portion of the weights of these layers on digital units. The reason is that these weights are usually more sensitive to noise. 
Interestingly, the number of weights in the first and last layers usually is not considerable compared to the other layers. For instance, for ResNet18 over CIFAR10, the number of parameters in the first and last layers are 1728 and 5120 respectively while for the seventeenth layer the number of parameters is 2359296. Even recent studies show that the last linear layers can be compressed significantly (e.g., 49$\times$ for VGG16)~\cite{han2015deep, han2016deep}. We observed that if we dedicate the first and the third last tiles to digital cores, it leads to a better performance and energy consumption.

Another feature of the proposed weight selection method is that the deviation among the percentage of protected weights per layer (except first and last layers that have dedicated tiles) in HybridAC is lesser than IWS method. For example, in the case of ResNet18, the standard deviation of percentage of protected weights per layer is 4.8$\times$ less than individual weight selection (1.37 vs 6.69). We observed the same trend for all the evaluated DNNs over different datasets. That means that the proposed approach provides more uniform weight selection distribution rather than the IWS~\cite{Dash2020, dash2021robust}. This property simplifies the hardware of HybridAC and increases resource utilization.

Figure \ref{figure:DigitalUnit} demonstrates the building blocks of the digital accelerator. Unlike conventional accelerators like Eyeriss that put PEs one one side and large buffer/memories on another side of the chip, we distribute tiny SRAMs throughout the chip besides computation units.  
The proposed architecture is inspired by WAX design~\cite{gudaparthi2019wire}. In each tile, we employ a tuple of tiny SRAM, a MAC unit, and three registers including activation register, weight register, and partial sum register.
In HybridAC, we utilize short interconnections, where it connects small SRAMs. These small SRAMs have been distributed among tiles and bring high data reuse of both weight and activation.
 
In HybridAC, we replace 54KB global SRAM buffer of Eyeriss~\cite{chen2016eyeriss} (considering 8-bit weights) or 6KB SRAM buffer of WAX~\cite{gudaparthi2019wire} with 1KB buffer
access (5.2$\times$ energy reduction). Like WAX architecture~\cite{gudaparthi2019wire}, we replace 12- and 24-entry register files and 224-byte scratchpad in Eyeriss with a single register access.
Unlike WAX, these units are connected through a grid structure rather than H-tree. This brings the advantages of more bandwidth, less area, and better energy efficiency as demonstrated in the previous works~\cite{shao2019simba, chen2019eyeriss}. 
Using H-tree makes the distance between these two neighbors as bad as $log(chip width)$. Although far tiles can communicate together thorough the grid structure, each tile usually needs to access each local SRAM or its neighbors (e.g., unlike WAX) or large, far-distant SRAMs (e.g., unlike Eyeriss-like designs). WAX architecture also needs additional muxing at each split point of its H-tree~\cite{rashidi2021enabling}. These hierarchical muxing is essential so that data from a neighboring sub-array is guided either to the other adjacent sub-array or to a central complex controller. This complex controller unit is eliminated in our proposed architecture. 

Additionally, the number of required units is almost 20\% of WAX because small portions of the weights are mapped to digital units. 

The digital part of these tiles are located face to face to reduce wire length communication. Besides, unlike WAX, each SRAM has 1 row for activations, 24 rows for weights, and 7 rows for partial sums (6X size reduction).

\subsection{Mapping Mechanism}
\label{sec:Map}
The hierarchical structure of the HybridAC enables to reduce data movement between different sub-units, having fewer wordlines and bitlines, and having higher internal bandwidth. In HybridAC one or several tile(s) are programmed to store the weights of each layer, where the input channels of a layer are mapped to the row of crossbars. The tiles are connected to each other in a pipelined manner.
Except for the first tile and the last three tiles, which are dedicated to digital accelerators, the remaining tiles have both digital and analog units. In case that one tile cannot accommodate the whole weights of a layer, the remaining are placed in the tile next to it. 

\begin{figure}[t!]
	\begin{center}
		\includegraphics[width =1\columnwidth]{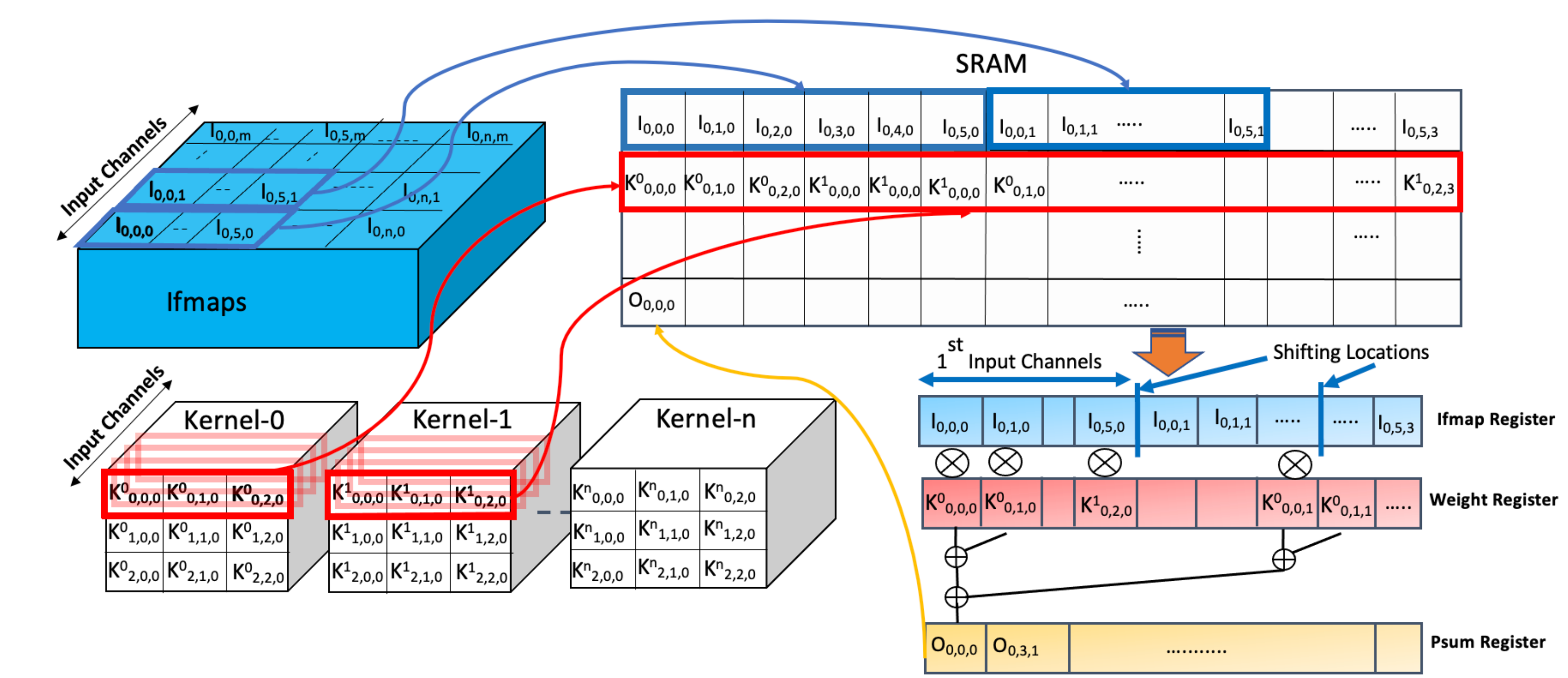}
	\vspace{-4ex}
	\caption{Digital flow of the proposed digital accelerator.
		\label{figure:dataflow}}
			\vspace{-2ex}
	\end{center}
	\vspace{-4ex}
\end{figure}

For the digital accelerator, as Figure~\ref{figure:dataflow} shows, the SRAM is 24-bit wide. The first row of the SRAMs is filled with the input feature maps of different input channels (i.e., 6 consecutive inputs of 4 different input channels). 
A row of weights from the first two kernels is placed in the first row dedicated for the weights in SRAM (e.g., second row). The remaining 23 weight rows of the SRAM are filled using the weights of the same position of other input channels. (e.g., (3 successive weights of 4 input channels for kernel1 and the same weights for kernel2).  
Once the inputs and weights are loaded from SRAM to registers and computation is started, loading of the next input feature maps is started such that we can have an overlap between computation and communication. When the weights are loaded into the SRAM subarray, they remain there until all weights are fully exploited. For activations, the subarray
is used to buffer the next row of activations.
Since we load weights and activations from different input channels (i.e. four different input channels), the corresponding activation and weights registers are split into 4 parts where shifting for each channel happens only in its corresponding partition. This multiplications happens using MAC units by multiplying weights and their corresponding activation. In the first cycle, 24 multiplications of corresponding weights and activation are performed. Each partial sum output is computed using adding four different outputs of multiplications using a 3-level adder tree. In 12 cycles, the 24 partial sum register are populated. Then, the results are written into the first row of SRAM dedicated to partial sum results (e.g., row 26). The same flow continues for the remaining activations and weights. Finally, results of the digital and analog portion for each layer are merged in the output register of each tile. We discuss more about the load balancing of HybridAC in Section \ref{sec:efficiency}.

\section{Simulation Setup}
\label{setup}
We evaluated five different DNNs, VGG16, ResNet18, ResNet34, DensNet121, and EfficientNetB3 over CIFAR10, CIFAR100, and ImageNet datasets. For all datasets, we compute five eigen pairs (values and vectors) as it has been demonstrated to deliver negligible accuracy drop~\cite{Dash2020, dash2021robust}. Simulation are performed in a hardware setup consisting of Intel 10$^{th}$ gen core i10 CPU and Nvidia GTX-1080Ti GPU.
\begin{figure}[b!]
	\vspace{-3ex}
	\begin{center}
		\includegraphics[width =1\columnwidth]{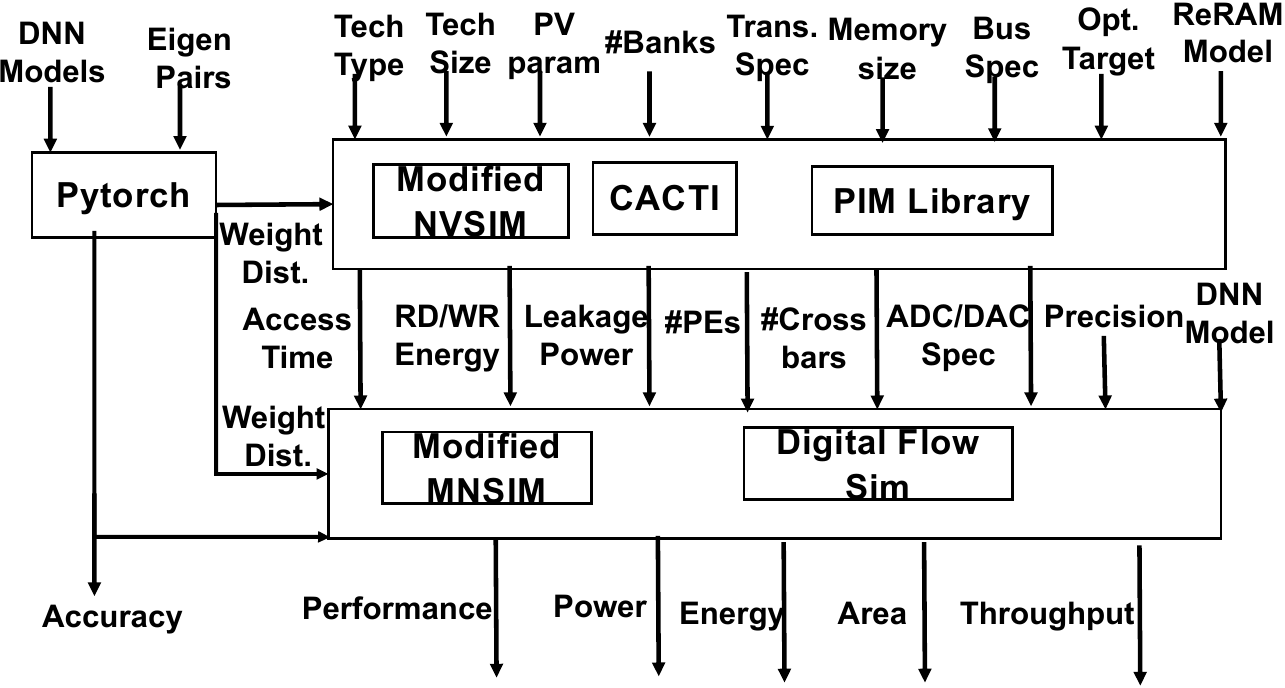}
	\vspace{-4ex}
	\caption{Overall flow of the developed in-house tool.
		\label{figure:Tool}}
	\end{center}
	\vspace{-4ex}
\end{figure}

We developed an in-house simulator to obtain the area, power, energy, and throughput of the HybridAC and evaluated baselines. Figure~\ref{figure:Tool} shows the overall flow of the developed simulator. 
It uses Pytorch API to get the accuracy and weight distribution as the algorithm-side is implemented in  Pytorch. The tool utilizes NVSIM~\cite{dong2012nvsim}, Cacti~\cite{balasubramonian2017cacti}, and PIM primitives library~\cite{wu2020architecture} as backbone simulators to perform design space exploration and gets required inputs of the architectural simulator. For the architectural simulator, we modify MNSIM\cite{zhu2020mnsim} to model the proposed hardware of analog units and also evaluated baselines. The tool is enriched with a digital simulator that mimics the proposed data flow.

Previous works~\cite{Long2019, Dash2020} suggest that conductance variation of the individual devices in a PIM design can be modeled by a Gaussian distribution of noise with $0$ mean and a standard deviation proportional to the stored parameter value. Eq. \eqref{eq:noisemodel} shows the noise model considered in our work:

\begin{equation}\label{eq:noisemodel}
{noise_{model}}
x_{i} \sim \mathcal{N}\left(0, \sigma w_{i}\right).
\end{equation}
We consider $\sigma=50\%$ and $\sigma=10\%$ for the weights in analog and digital cores, respectively.
We employ the VTEAM ReRAM model~\cite{kvatinsky2015vteam} and follow the methodology of ISAAC to model max-pooling, shift-and-add, ADC, and activation functions. The power and area of the shift-and-add, max-pool, and ReLU are taken from ISAAC~\cite{shafiee2016isaac} and PRIME~\cite{PRIME}. 
 
We employ HyperTransport serial links the same as the one used by ISAAC~\cite{shafiee2016isaac} and DaDianNao~\cite{chen2014dadiannao}.

To get the power and area of ADC, we use the most updated dataset~\cite{murmannadc}. To extract the area and power of the same ADC but with a lower resolution, we scale down the power and area of the memory, clock, and \textit{vref} buffer linearly, and the capacitive DAC exponentially as explained in~\cite{saberi2011analysis}. 
We choose this methodology to model peripheral circuitry and make a fair comparison at 32nm with state-of-the-art works. We also compare the results with the fully digital SIGMA~\cite{qin2020sigma}.
We examine both offset subtraction-based designs~\cite{shafiee2016isaac, RSE2019, ankit2019puma}, (i.e., HybridAC, HybAC) and differential cell-based designs ~\cite{PRIME, yao2020fully, joshi2020accurate}( i.e., HybridACDi, HybACDi) to show HybridAC mechanism is independent to the underlying architecture.
 
 We implemented the RTL design of the proposed digital units and the baselines using Verilog-HDL that enables us to compute the power and area more accurately.
 We utilize Synopsys Design Compiler at 28nm, a Low Leakage library, and a clock frequency of 1GHz to synthesize the digital design. To get SRAM results we utilize Synopsys memory compiler.  We employ Innovus to conduct the layout for the digital part, where the extracted SRAM using memory compiler is used in the layout as a hard macro. The results are scaled up to 32nm. The extracted results are back annotated in the developed simulator to help us to get more accurate numbers.

\section{Result}
\label{sec:results}

\subsection{Accuracy vs Protected Weight Percentage}
\begin{figure*}[!ht]
	\hspace{0.5cm}
	\begin{minipage}[b]{.30\linewidth}
		\centering
		\includegraphics[width =1\columnwidth]{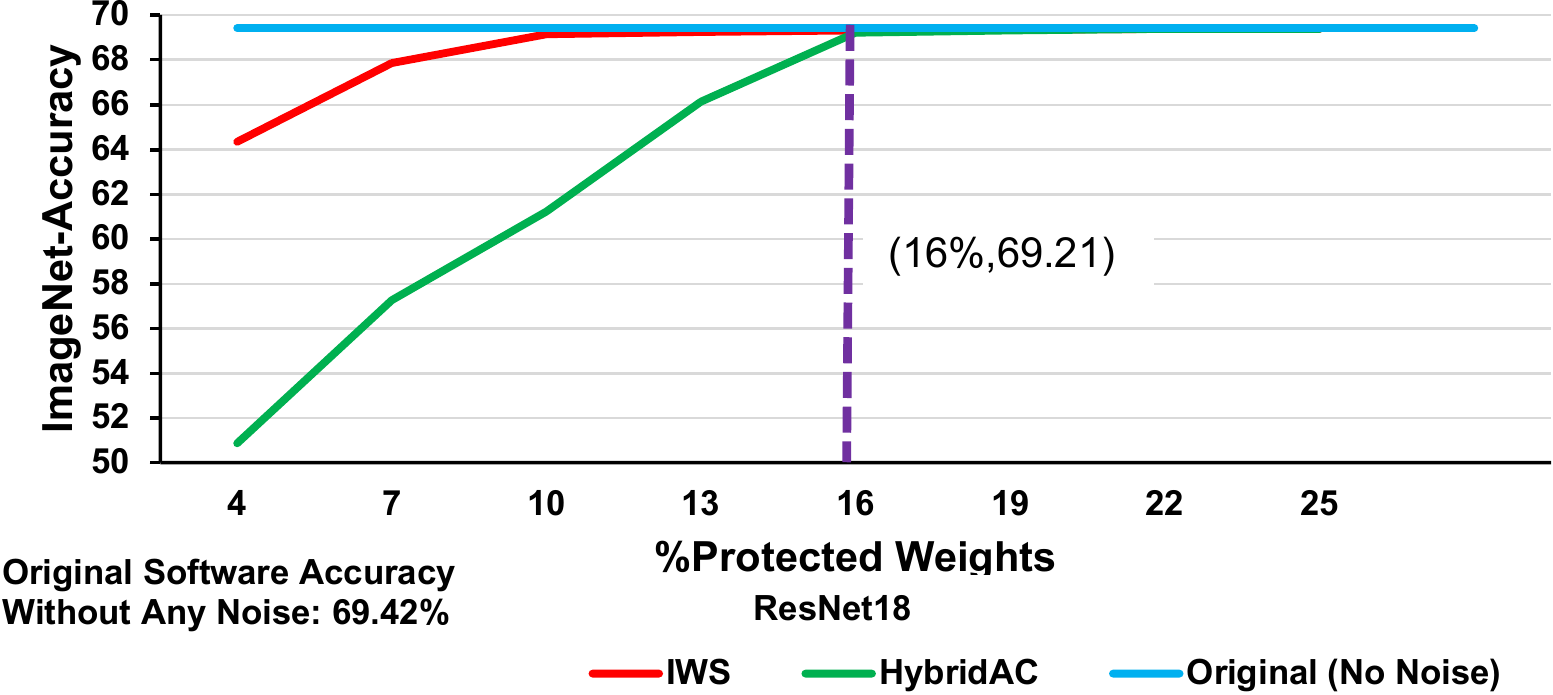}
	\end{minipage}
	\begin{minipage}[b]{.30\linewidth}
		\centering
		\includegraphics[width =1\columnwidth]{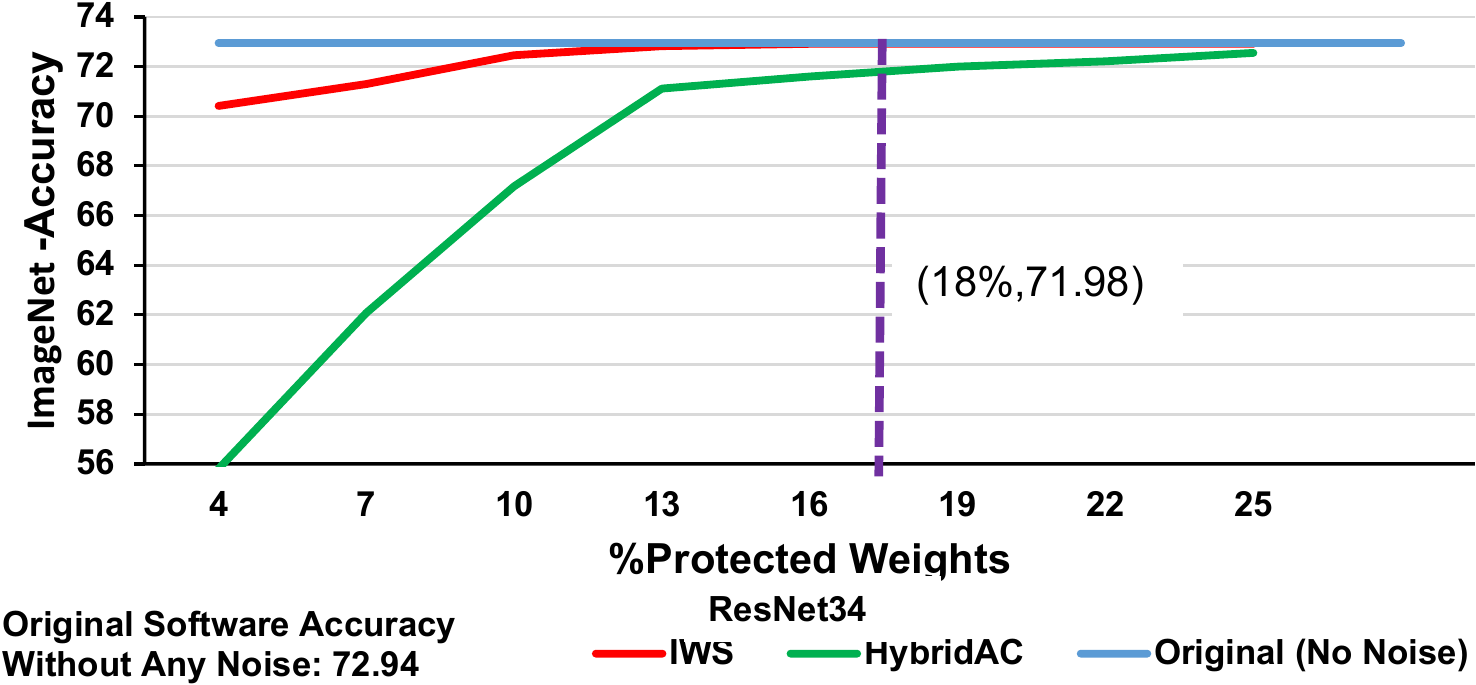}
	\end{minipage}
	\begin{minipage}[b]{.30\linewidth}
		\centering
		\includegraphics[width =0.95\columnwidth]{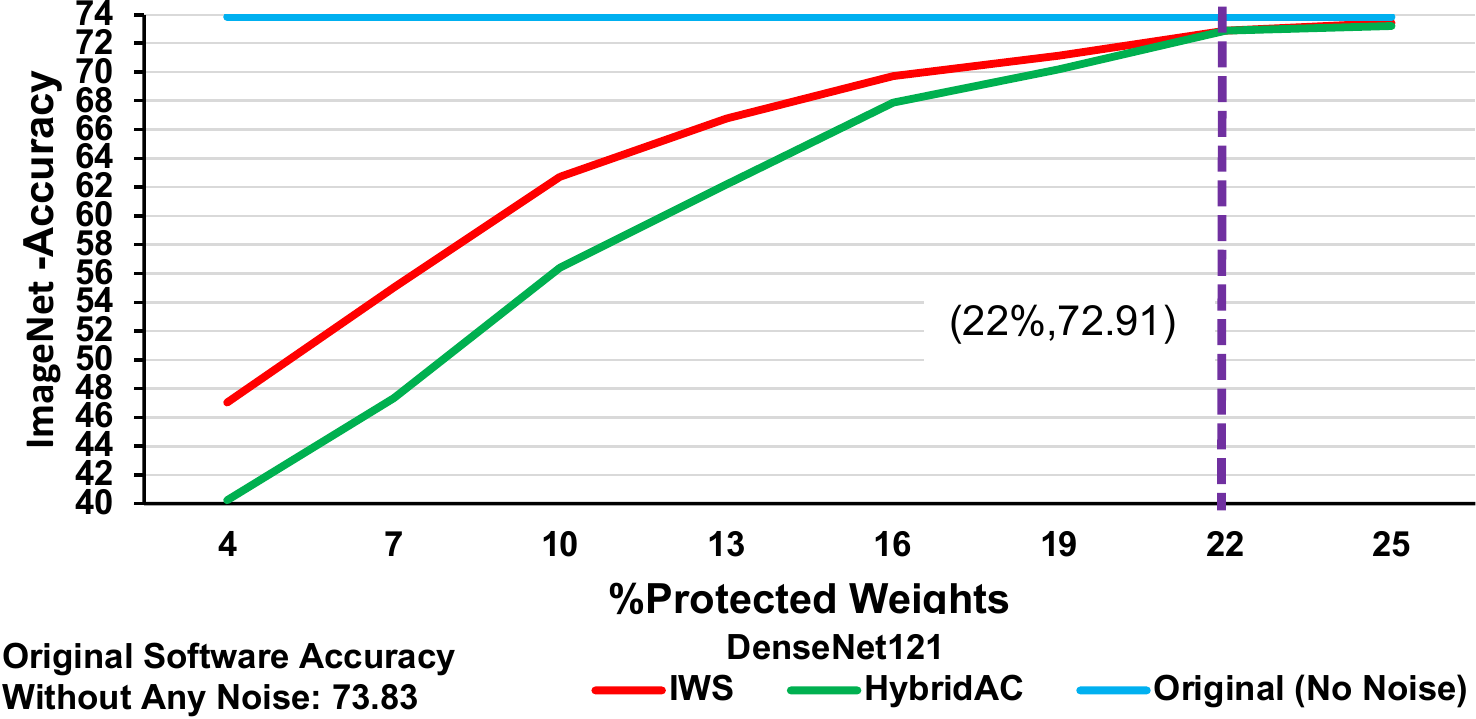}
		\label{figure:Prot_ChW_ImageNet_DensNet}
	\end{minipage}
	\hspace{0.5cm}
 	\vspace{-2ex}
	\caption{Comparing the impact of IWS and HybridAC approaches on accuracy over ImageNet dataset.
	\label{figure:Prot_ChW_IamgeNet_ResNet18ResNet34DenseNet121}}
\vspace{-3ex}
\end{figure*}

 Table~\ref{table:Prot_ChW_CIFAR} illustrates the accuracy results of CIFAR10/100 dataset of the proposed approach and compare it with the IWS method~\cite{Dash2020,dash2021robust}. We repeat the experiment 50 times and get the average of the results. The third column shows the accuracy when there is no variation, The fourth column shows the results when there are $50\%$ and $10\%$ variations for the weights placed in the analog accelerator. The 
 ``\%Selected Weights" column shows the percentage of the weights that need to be placed in digital cores to reach the desirable accuracy, which is typically less than 1\% of the original accuracy~\cite{darvish2020pushing}. Plots~\ref{figure:Prot_ChW_IamgeNet_ResNet18ResNet34DenseNet121} plots demonstrates the results along with its trend over ImageNet dataset. The x axis shows the protected weight percentage and the y-axis shows the accuracy. For ImageNet, the accuracy without any protection mechanism is 7.25\%, 17.13\%, and 4.4\% for ResNet18, ResNet34, and DenseNet121, respectively.
 
 As the results illustrate,  the amount of the weights that needs to be protected to reach the desired accuracy varies depending on the DNNs, dataset, number of layers, parameters, and most importantly weight distributions.
 In general, when the DNN and dataset become more complex, the amount of protected weight is increased. For example, the percentage of protected weight for Imagenet is more than CIFAR100 and for CIFAR100 is more than CIFAR10 for the same DNN. On the other hand, for the same dataset, in general, DenseNet121 needs the highest weight percentage to be protected among evaluated networks. 
 
 Compared with the IWS, the HybridAC pushes more weights to the digital units since it selects the whole channel rather than choosing the individual weights.
 Fortunately, this mechanism leads to a better load balancing and higher hardware utilization. We discuss this feature in Section~\ref{sec:efficiency}.

\begin{table*}[ht!]
	
	\vspace{2ex}
	\caption{Comparing the impact of IWS and HybridAC approaches on accuracy over CIFAR10/100 datasets \label{table:Prot_ChW_CIFAR}}
	\centering{\scriptsize\setlength\tabcolsep{6pt}
		\vspace{-2ex}
		\begin{tabular}{|l|c|c|c|c|c|c|c|}
			\hline	&\multicolumn{6}{|c|}{\textbf{HybridAC/IWS~\cite{Dash2020,dash2021robust}}}\\ 
			\hline
			\textbf{DataSet}
			&
			\textbf{DNN} 
			& 
			\textbf{Original Accuracy} &
			\textbf{Accuracy with PV} &
			\textbf{\%Selected Weights}&\textbf{Accuracy}& \textbf{\%Selected Weights} & \textbf{Accuracy}\\ 
			&& \textbf{with No Noise} & \textbf{Both}&\textbf{IWS~\cite{Dash2020}}& \textbf{\textbf{IWS~\cite{Dash2020}}} & \textbf{HybridAC}& \textbf{HybridAC}\\
			\hline
            \hline
            &
			VGG16&  92.01\% &27.53\%& 4.00\% & 91.81\%& 10.00\% &91.69\%\\ 
			\hline
			&
			ResNet18& 94.76\%  &25.66\%& 3.00\% &94.53\%& 10.00\% &94.26\%\\
			\hline 
			CIFAR10&
			ResNet34  & 93.93\% &55.01\%& 6.00\% &92.98\%& 9.00\% &92.94\%\\
			\hline 
			&
			DenseNet121  & 94.96\% &22.77\%& 6.00\%& 94.32\%& 11.00\% &94.04\%\\
			\hline 
			&
			EfficientNetB3  & 95.22\% &23.88\% & 5.00\%&94.46\% &10.00\%&94.36\%\\
			\hline
			\hline
			&
			VGG16&  69.29\% & 12.3\%& 4.00\%  &68.35\%& 12.00\%  &68.30\%\\
			\hline
			&
			ResNet18& 74.07\% &13.86\%& 8.00\%& 73.73\%&13.00\% &73.51\%\\
			\hline 
			CIFAR100&
			ResNet34 & 74.78\% &24.28\%& 8.00\% &74.06\%& 14.00\%&73.99\%\\
			\hline 
			&
			DenseNet121  & 73.9\% &10.16\%& 10.00\%&73.12\% &16.00\%&72.80\%\\
			\hline 
			&
			EfficientNetB3  & 82.45\% &13.20\%& 8.00\%&81.61\% &14.00\%&81.50\%\\
			\hline
		\end{tabular}
	\vspace{-2ex}
	}

\end{table*}


\subsection{ADC Resolution}
 
In theory, the required bits for a full ADC resolution is computed using the equation~\ref{eqn:ADCresolution}~\cite{shafiee2016isaac}.
 \begin{equation}
    ADC_{bits}=
    \begin{cases}
      v + w + log(r), & \text{if}\ v>1\ \&\  w>1 \\
      v + w + log(r) -1, & \text{otherwise}
    \end{cases}
    \label{eqn:ADCresolution}
  \end{equation}

Where \textit{v} denotes number of bits per input, \textit{w} shows weight bits stored per ReRAM cell on the crossbar arrays, and \textit{r} represents activated wordlines per crossbar array~\cite{shafiee2016isaac}. For the evaluated architecture, we consider 2 bits per cell while 128 rows can be activated at the same time and the number of bits per input is 1. We employ the same encoding technique used in ISAAC to save 1-bit ADC. Accordingly, for the mentioned configuration, the required ADC resolution with zero accuracy loss assuming there is no source of variation is 8 bits.  

Table~\ref{table:ADCres} shows accuracy results when ADCs with different resolutions are used for HybridAC with bias and offset subtraction architecture (i.e., HybAC), baseline(i.e., IWS~\cite{Dash2020})), and also differential cell architecture (i.e., HybACDi, IWSDi). 

As the results illustrate, for 7-bit ADC, in the case of CIFAR10/100 the accuracy results of IWS is slightly higher than HybridAC because its original accuracy is higher. However, by reducing ADC resolution, the gains of the proposed solution become more apparent. For instance, in the worst case, 0.29\%, 0.45\%, and 0.66\% accuracy drop happens for CIFAR10, CIFAR100, and Imagenet datesets, respectively. Nonetheless, using IWS method~\cite{Dash2020}, the accuracy drops are 0.98\%, 1.95\%, and 3.87\%, respectively, which, however comes at the cost of using higher bit-resolution (i.e. 8-bit) ADC.

In the proposed design, the channels are mapped to rows. Hence, by transferring a channel, the whole row values are moved to the digital accelerator. ADCs are defined per bitlines. Removing (a) row(s) will have an uniform impact on whole bitlines of a crossbar and hence ADC resolution. For IWS method ~\cite{Dash2020, dash2021robust}, we are not able to have the same ADC for all bitlines. The reason is that important weights have a random distribution. As a result, the number of important weights may change dramatically by going from one bitline to another one. 
Furthermore, although only small portion of channels are transferred to the digital accelerator, they have a greater impact on the final accuracy. Removing these critical weights enables us to employ ADCs with smaller resolution and less less area/power while having minimal accuracy drop.

In the last two columns, we assess more extreme situations where we use 4-bit ADCs and reach almost the same accuracy as 6-bit ADC but through different architecture. The reason is that, here, we examine differential cells~\cite{PRIME, yao2020fully, joshi2020accurate} where two separated crossbars are used for matrices
with positive and negative weights. Unlike bias and offset subtraction cells~\cite{shafiee2016isaac, RSE2019, ankit2019puma}, this mapping does not add biases to the weights, and accordingly their corresponding conductance are reduced. This is helpful as in many  DNNs, zero and low-value weights contribute significantly in distribution of weight values~\cite{zhang2018systematic,zhao2019improving,han2021improving}. Accordingly, the conductance (weight) variation and parasitic voltage drop across the crossbar columns (rows) are decreased, that results in a higher accuracy. This fact shows that the proposed method is independent of the underlying architecture. It is noteworthy to mention that differential cell-based design have more hardware overhead and generally less throughput than offset-based designs.

Reducing ADC resolution is important as ADC contributes more than 30\% of tile area and 50\% of tile power in many architecture~\cite{shafiee2016isaac}. Having 7-bit ADC instead of 8-bit will save 7\% area and 14\% power of the tile. In the case of 6-bit ADC, the saving area and power of the tile would be 13\% and 29\%. By employing 6-bit ADC and pushing a percentage of weights to digital units as shown in the Table~\ref{table:Prot_ChW_CIFAR} 
and Figure~\ref{figure:Prot_ChW_IamgeNet_ResNet18ResNet34DenseNet121} with the acceptable accuracy drop, the power/area saving is considerable. Obviously, by pushing more weights to digital accelerator, we can increase the accuracy at the cost of having lower throughput and performance.

\subsection{Hybrid Quantization}
Table~\ref{table:quant} demonstrates the impact of hybrid quantization along with various ADC resolutions on the final accuracy. The second column shows the impact of hybrid quantization where important weights in digital cores are 8-bit and less important weights placed in ReRAM crossbars are 6-bit. As seen in the best and the worst case scenarios, the accuracy drops by 0.09\% (i.e., ResNet18/ CIFAR10), and 0.53\% (i.e., DenseNet121/ ImageNet), respectively, compared to the situation where all weights are 8 bits. The last column shows the impact of hybrid quantization along with employing 6-bit ADC together. As shown, in the best case (ResNet18/ CIFAR10) 0.30\% and in the worst case (DenseNet121/ ImageNet) 1.11\% accuracy loss happens, compared to the situation when there is no smaller ADC and hybrid quantization. 

As we mentioned in Section~\ref{sec:quant}, mapping important channels in the digital and others in the analog units enables us to have the first work  that proposes a mixed-precision quantization schemes for different input channels of a single layer without any need for retraining. Obviously, by employing post-quantization fine-tuning, we can even achieve smaller quantization bits~\cite{deng2020model, mckinstry2019discovering}.
 
\begin{table*}[t!]
	\begin{minipage}[b]{.7\linewidth}
	\vspace{2ex}
	\caption{ Comparing impact of using ADC with smaller resolution on accuracy of both HybridAC and IWS~\cite{Dash2020}\label{table:ADCres}.}
	\centering{\scriptsize\setlength\tabcolsep{6pt}
		\vspace{-2ex}
		\begin{tabular}{|l|c|c|c|c|c|c|c|c|c|}
			\hline
			&\multicolumn{9}{|c|}{\textbf{HybridAC/IWS~\cite{Dash2020,dash2021robust}}}\\ 
			\hline
			\textbf{DSet}& 
			\textbf{DNN}&
			\textbf{8-bit} & \textbf{8-bit}& \textbf{7-bit}&\textbf{7-bit}& \textbf{6-bit} & \textbf{6-bit}&\textbf{4-bit}&\textbf{4-bit}\\
		    \hline
		    &
			&\textbf{HybAC}  &\textbf{ IWS}& \textbf{ HybAC}&\textbf{IWS}& \textbf{HybAC} & \textbf{IWS} & \textbf{HACDi} & \textbf{IWSDi}\\
			\hline
			\hline
            &
            
            
			VGG16 &91.69\% & 91.81\% & 91.57\% &91.60\%& 91.45\% &91.15\%& 91.39\% &91.05\%\\ 
			\hline
			&
			ResNet18 &94.26\% & 94.53\% & 94.20\% &94.31\%& 94.05\% &93.89\%& 93.98\% &93.82\%\\
			\hline 
 			\textbf{C10}&
			ResNet34 &92.94\% & 92.98\% & 92.90\% &92.88\%& 92.77\% &92.47\%&92.68\% &92.41\%\\
			\hline 
			&
			DenseNet121&94.04\% & 94.32\% & 93.89\%& 94.09\%& 93.75\% &93.34\%&93.62\% &93.29\%\\
			\hline 
			&
			EfficientNetB3  & 94.36\% & 94.46\% & 94.26\%& 94.39\%& 94.12\% &93.66\%&94.01\% &93.61\%\\
			\hline
			\hline 
			&
			VGG16 &68.30\%&  68.35\% & 68.11\%  &68.01\%& 67.98\%  &66.66\%&67.88\%  &66.57\%\\
			\hline
			&
			ResNet18&73.51\%& 73.73\% & 73.40\%& 73.43\%&73.23\% &72.17\%&73.17\% &72.09\%\\
			\hline 
			\textbf{C100}&
			ResNet34&73.99\%& 74.06\% & 73.85\% &73.85\%& 73.60\%&72.54\%&73.52\%&72.48\%\\
			\hline 
			&
			DenseNet121& 72.80\%& 73.12\% & 72.69\%&72.75\% &72.35\%&71.17\%&72.29\%&71.06\%\\
			\hline 
			&
			EfficientNetB3  &81.50\%& 81.61\% & 81.37\%& 81.42\%& 81.25\%& 81.08\%&81.13\% &80.95\%\\
			\hline
			\hline 
			&
			ResNet18 &69.21\% & 69.21\% & 68.99\% &67.55\%& 68.65\%&66.08\%&68.56\%&65.95\%\\
			\hline 
			\textbf{IN}&
			ResNet34& 71.98\% & 72.20\%&71.84 &70.06\%& 71.53\%&68.95\%&71.42\%&68.81\%\\
			\hline 
			&
			DenseNet121 &72.91\% & 72.91\% & 72.61\%&72.28\% &72.25\%&69.04\%&72.12\%&68.92\%\\
			\hline
		\end{tabular}
	\vspace{-2ex}
	}
	\end{minipage}
	\hspace{1ex}
	\begin{minipage}[b]{.25\linewidth}
    \vspace{2ex}
	\caption{ Comparing impact of hybrid quantization on accuracy. \label{table:quant}}
	\centering{\scriptsize\setlength\tabcolsep{6pt}
		\vspace{-2ex}
		\begin{tabular}{|l|c|c|c|}
			\hline
		
			\textbf{DSet}& 
			\textbf{DNN}&\textbf{(8-6)}&\textbf{(8-6)}\\ 
			& 
			&
		  \textbf{8-ADC}& \textbf{6-ADC} \\ 
			\hline
			\hline
			&
			VGG16 &   91.53\%  & 91.30\% \\ 
			\hline
			&
			ResNet18 & 94.17\% & 93.96\% \\
			\hline 
			\textbf{C10}&
			ResNet34  & 92.83\% & 92.61\% \\
			\hline 
			&
			DenseNet121  & 93.89\% & 93.58\% \\
			\hline 
			&
			EfficientNetB3   & 94.29\%&93.97\%\\
			\hline
			\hline
			&
			VGG16   & 68.10\%  & 67.71\%  \\
			\hline
			&
			ResNet18 & 73.37\% &73.04\%\\
			\hline
			\textbf{C100}&
			ResNet34  & 73.78\% & 73.40\%\\
			\hline 
			&
			DenseNet121& 72.69\% &72.10\%\\
			\hline 
			&
			EfficientNetB3  & 81.46\%& 81.07\%\\
			\hline
			\hline
			&
			ResNet18  & 68.83\% & 68.29\%\\
			\hline 
			\textbf{IN}&
			ResNet34  & 71.49\% & 71.07\%\\
			\hline 
			&
			DenseNet121& 72.40\% &71.80\%\\
			\hline
		\end{tabular}
	\vspace{-2ex}
	}
	\end{minipage}
\end{table*}

\subsection{Comparison with More Related Works}

\subsubsection{Power and Area}
\label{sec:powerarea}
Table 5 shows the power and the area of different components of HybridAC and its comparison with ideal-ISAAC as a baseline architecture in details. Tables 6 and 7 summarize the numbers of Table 5 considering the required number of MCUs and Tiles in more coarse granularity. These tables also compare the power and area of HybridAC with assorted baselines.
As we discussed in the Section~\ref{sec:BBAnalog}, because of moving critical weights to the digital cores, HybridAC needs fewer crossbars, smaller peripheral circuitry such as eDRAM, ADC, and sample-and-hold logic. HybridAC needs bigger quantization circuitry as we explained in Section~\ref{sec:quant}. However, it requires only 8 MCUs per tile, while  other designs usually need 12 MCUs per tile. In addition the total number of required tiles reduces from 168 (in ISAAC) to 148. The bottom of the Table 5 provides the power and area of the digital accelerator of HybridAC.

As demonstrated, HybridAC provides 28\% better area and 57\% better power compared to ISAAC. For IWS~\cite{Dash2020, dash2021robust}, we consider two different approaches. In the first one (IWS-1), there is only 1 tile and the weights are written into crossbar cells after every layer is finished. Although this approach  saves the area and power of the analog unit, the performance and the throughput will drop significantly. The reason is that writing the weights into ReRAM cells takes a considerable amount of time (e.g., 50ns unipolar and 200ns bipolar). Moreover, we need to write into a cell multiple times to make sure it has a correct value~\cite{xu2015overcoming}. Using the same tile for writing the weights of all layers makes the endurance issue of ReRAM technology worse~\cite{mittal2019survey}. Besides, this method limits the parallelism to only one tile. 

Accordingly, most of the existing works~\cite{shafiee2016isaac, PRIME,li2020timely} will consider another approach where they write all (or most) weights once into the different crossbars in various tiles. IWS-2 follows this methodology. It impose weights overheads as they leave zeros in the crossbars in the place of the weight that are transferred to digital units. Those zeros (up to 22\% in the case of DenseNet121/ImageNet) that randomly are distributed among other weights add up to 400 more crossbars to existing ones, which increase area, energy and execution time.

HybridAC improves the power and the area of the IWS-2~\cite{Dash2020} by 65\% and 2.1$\times$ respectively. IWS-1 has a slightly better power (e.g., 1\%) and worse area than HybridAC (47\%) due to using costly SIGMA~\cite{qin2020sigma} that has high power and area. 

As a digital accelerator, SIGMA exploits a configurable systolic based architecture with a lot of interconnection overheads. Consequently, SIGMA has a relatively lower area efficiency (i.e., $155\frac{GOPS}{mm2}$).
In comparison with SIGMA, HybridAC provides 15\% better area. HybridAC outclasses FORMS~\cite{yuan2021forms} by 44\%(29\%) in terms of power and area, while the improvement over SRE~\cite{RSE2019} are 45\%(27\%). 
Although digital accelerators show better power consumption, the area and power efficiency of the HybridAC outperforms the existing accelerators as Table~\ref{table:GOPS} shows.

\begin{table}[b!]
\footnotesize
	\vspace{-3ex}
	\caption{Peak area- and power-efficiency of different architectures normalized to Ideal-ISAAC.}
	\label{table:GOPS}
	\centerline{
		\scalebox{1.1}{
	{\scriptsize\setlength\tabcolsep{9pt}
		\begin{tabular}{|c|c|c|}
			\hline
			\textbf{Architecture}  &\textbf{$\frac {GOPs}{s \times mm^{2}}$}  & \textbf{$\frac {GOPs}{s \times W}$} \\
			\hline
			\hline
			Ideal-ISAAC~\cite{shafiee2016isaac} & 1 & 1   \\
			PUMA~\cite{ankit2019puma} & 0.70 & 0.79   \\
			SRE\cite{RSE2019}&0.19&0.26 \\
			FORMS8(not pruned)\cite{yuan2021forms}&0.54&0.61\\
			FORMS16(not pruned)\cite{yuan2021forms}&0.77& 0.84\\	DaDianNao~\cite{chen2014dadiannao} & 0.13 & 0.45   \\
			TPU~\cite{tpu} &0.08 & 0.48 \\
			WAX\cite{gudaparthi2019wire}& 0.33 &2.3\\
			SIMBA\cite{shao2019simba}& 0.34-0.62 &0.08-2.4\\
            IWS-1\cite{Dash2020, dash2021robust}  & 0.13 & 0.15 \\
			IWS-2\cite{Dash2020, dash2021robust}  & 0.38 & 0.41 \\
			HybridAC & 1.43 & 1.81 \\
			HybridACDi & 1.75 & 2.5\\
			\hline
		\end{tabular}}
	}
}
\end{table}
\begin{figure}[b!]
	\vspace{-2ex}
	\begin{center}
		\includegraphics[width =1\columnwidth]{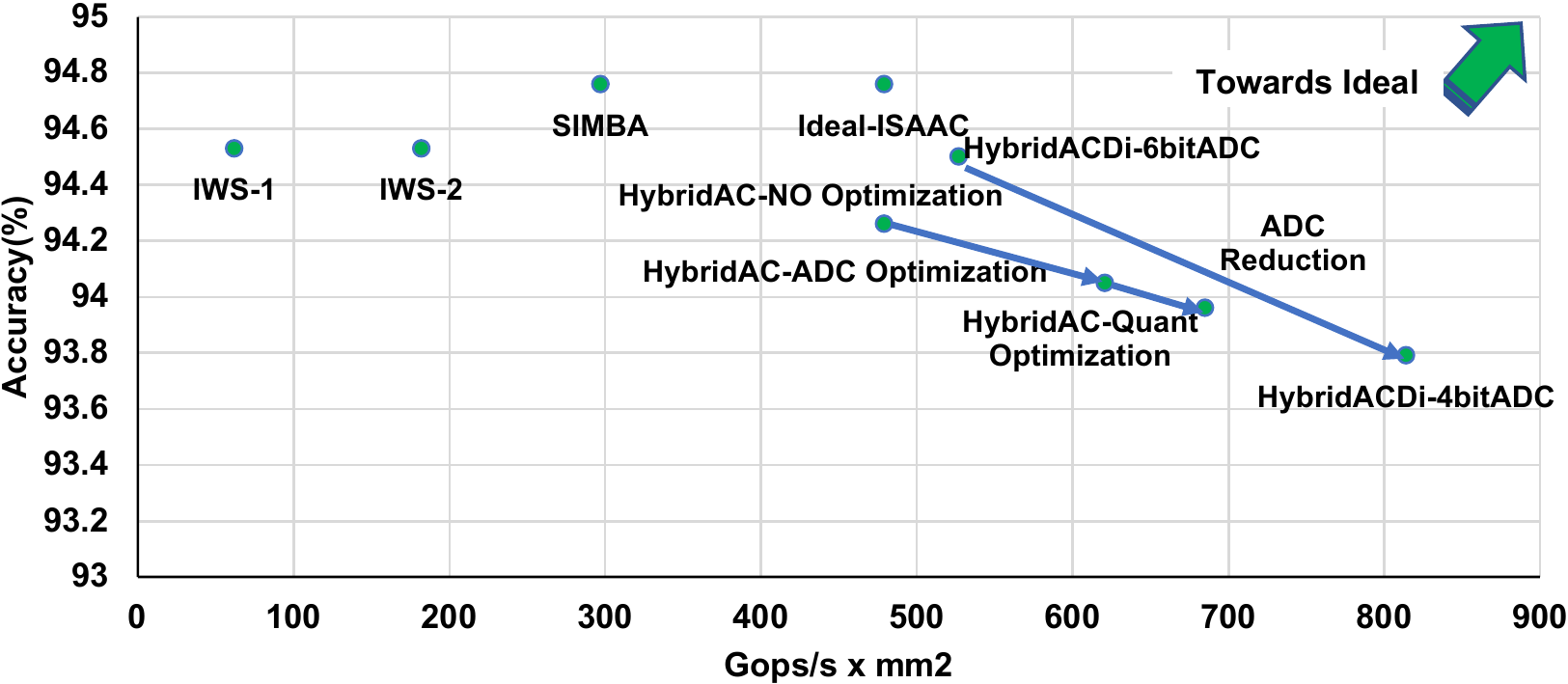}
	\vspace{-3ex}
	\caption{Accuracy vs area-efficiency comparison for different architectures for ResNet18 inference on CIFAR10.
		\label{figure:stands}}
	\end{center}
	\vspace{-6ex}
\end{figure}

\subsubsection{Comparison Summary: Area/Power Efficiency}
\label{sec:efficiency}
Table~\ref{table:GOPS} presents the area and power efficiency of HybridAC (e.g., using 6-bit ADC) and HybridACDi (using 4-bit ADC) and also assorted digital and analog accelerators. As seen, HybridAC and HybridACDi outperform all of them while they are more immune to variation. The efficiency of HybridAC comes from the novel design that utilizes smaller ADCs, hybrid quantization, fewer crossbars, and tiny but high throughput digital accelerator. The area- and power-efficiency of Ideal-ISAAC (e.g., it idealistically activates 128 wordlines considering it is immune to the noise while in practice it cannot reach that) in 32nm technology while the frequency is 1GHz and both weight and activation are 8 bits are 1912 $\frac{GOPS}{s \times mm^2}$ and 2510 $\frac{GOPS}{s \times w}$. 

Figure~\ref{figure:stands} shows the accuracy versus area-efficiency of different architectures for Inference of ResNet18 over CIFAR10. As it can be observed, it clearly shows how the application of different optimizations on HybridAC can help us to approach more to the ideal point.

In the case of HybridAC, the digital and analog accelerators can have different clocks. However, to make the design simpler we use the same clock frequency.
We consider that the frequency of both digital and analog as 1GHz while both of the designs are built in 32nm technology. Our results show that the peak area-efficiency of the analog accelerator is 2549 $\frac{GOPS}{s \times mm^2}$. However, the throughput of the digital cores are 434 $\frac{GOPS}{s \times mm^2}$. 
That means that the area-efficiency of analog is $\frac{2549}{434}$ is 5.87$\times$ bigger than digital one. This implies that for an ideal load balanced scenario such that digital and analog ones finish at the same time, the analog one needs to do 5.87$\times$ more work than the digital. For instance, almost 16\% of weights need to be handled by the digital accelerator. This is reachable for all of the DNNs except DenseNet121 over ImageNet. Note that the results in Table~\ref{table:Prot_ChW_CIFAR} and Figure~\ref{figure:Prot_ChW_IamgeNet_ResNet18ResNet34DenseNet121} are the minimum percentage of weight that need to be pushed in digital to reach the high accuracy (i.e., accuracy loss is less than 1\% of the ideal one considered in many applications~\cite{darvish2020pushing}).

\begin{figure*}[ht!]
	\vspace{-3ex}
	\begin{center}
	
	{Table 5: Power and area of different components of HybridAC and its comparison with ISAAC}
	{table:ASICresultcom}{}
	\includegraphics[width = 2\columnwidth]{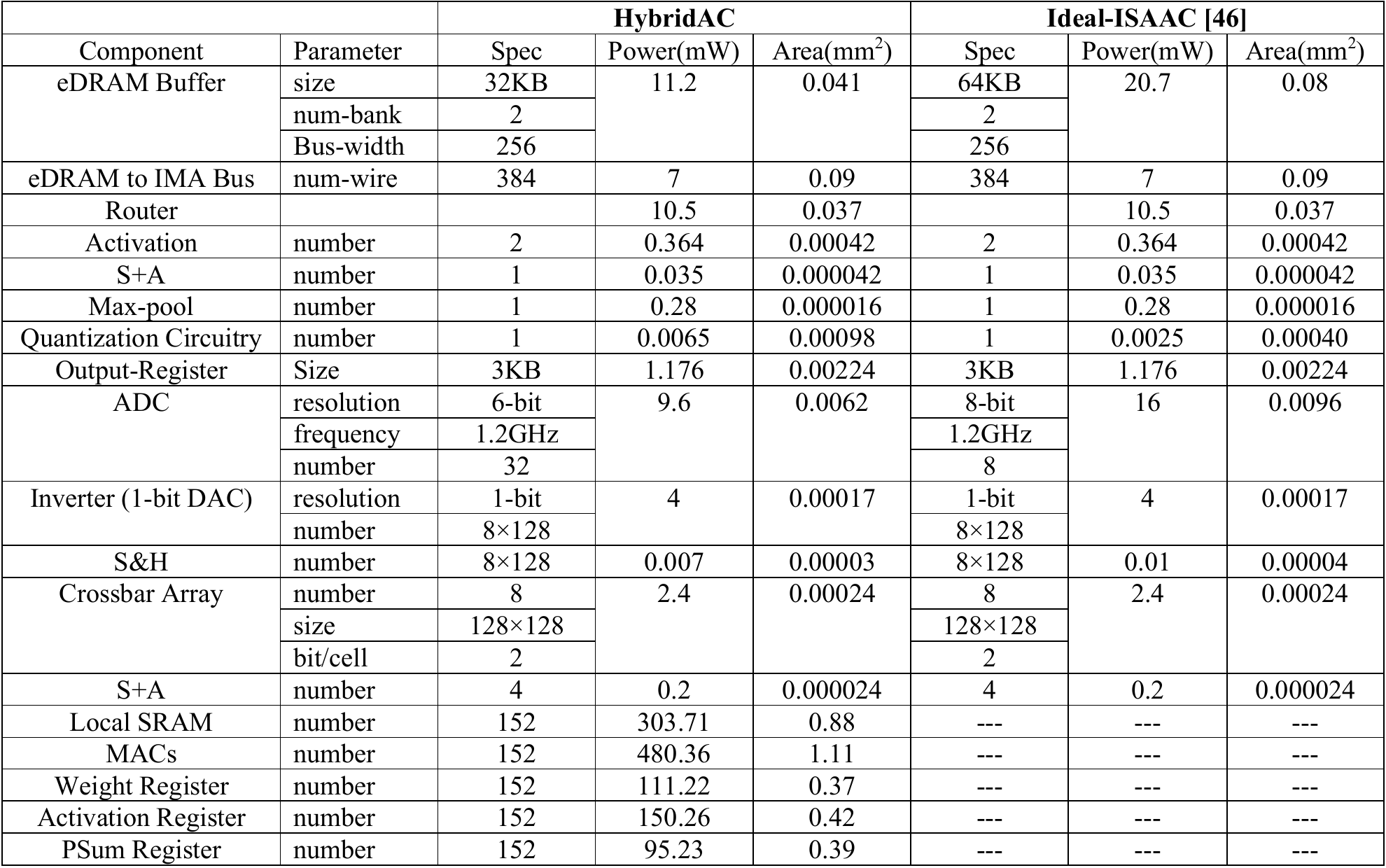}\\
	\end{center}
\end{figure*}

\begin{figure*}[ht!]
	\vspace{-3ex}
	\begin{center}
	{Table 6: Comparison of total power and area of HybridAC with different baselines}
	{table:ASICresultsummary1}{}
	\includegraphics[width = 2.1\columnwidth]{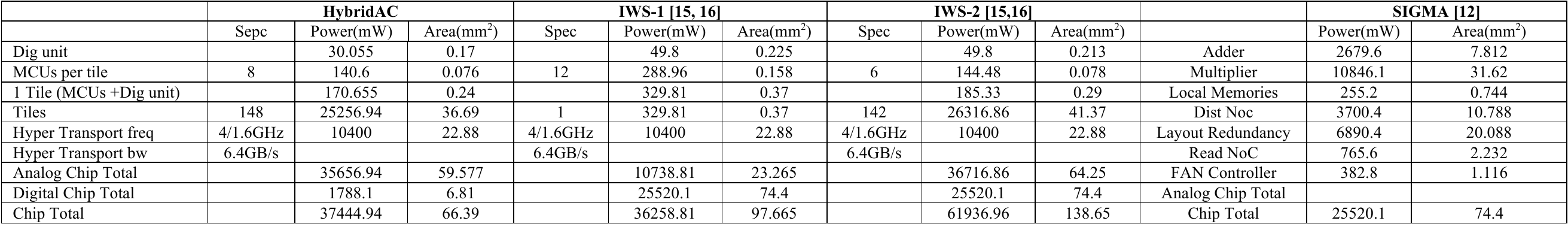}\\
	\end{center}
\end{figure*}

\begin{figure*}[ht!]
	\vspace{-3ex}
	\begin{center}
		
	{Table 7: Comparison of total power and area of HybridAC with different baselines-Continued}
		{table:ASICresultsummary2}{}
	\includegraphics[width = 2.1\columnwidth]{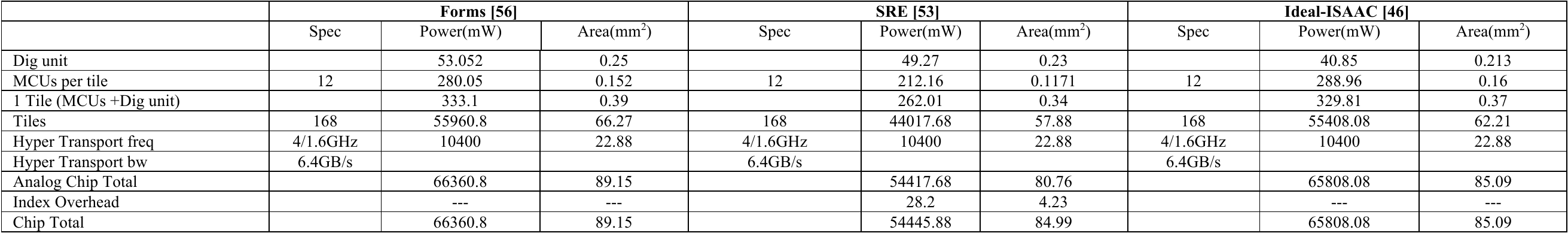}\\
	\end{center}
\end{figure*}

\subsubsection{Execution Time and Energy Results}
\begin{figure}[ht!]
	\vspace{-3ex}
	\begin{center}
		\includegraphics[width =1\columnwidth]{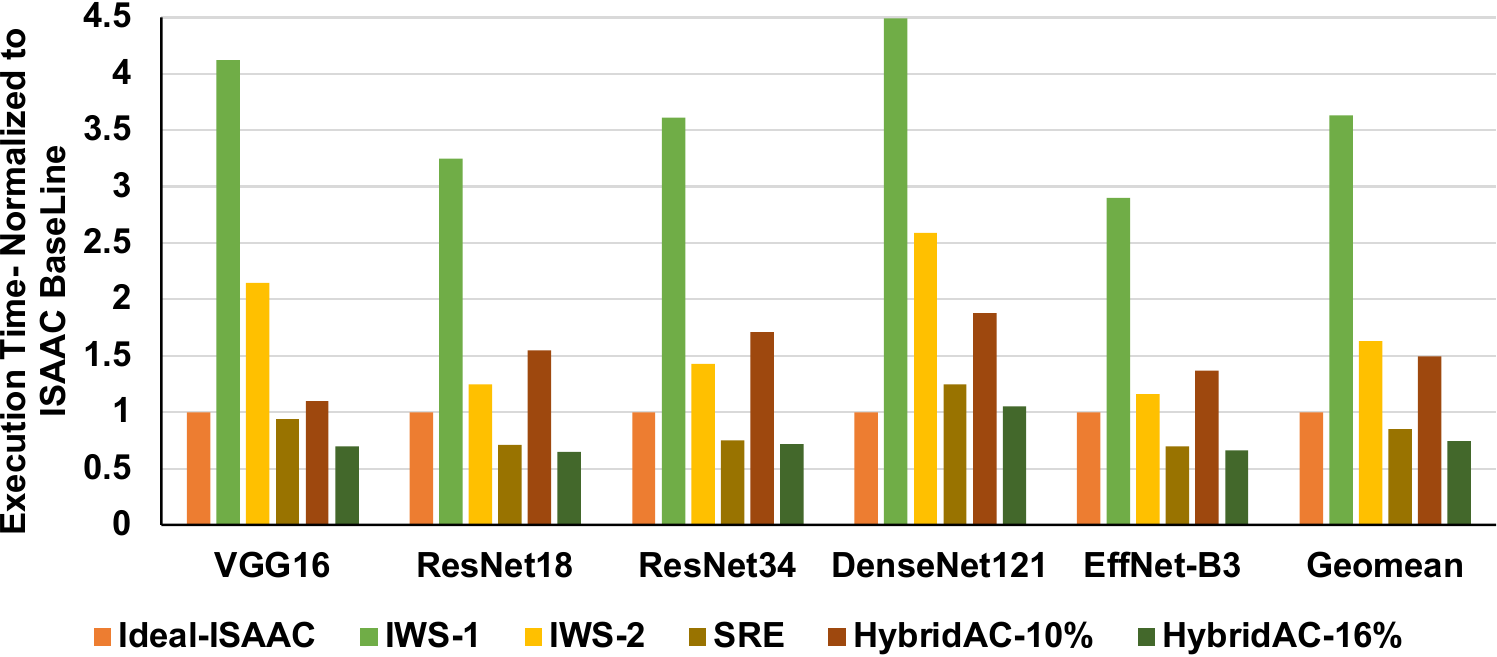}
	\vspace{-4ex}
	\caption{Execution time of different DNNs using different architectures
		\label{figure:time}}
	\end{center}
	\vspace{-2ex}
\end{figure}

\begin{figure}[ht!]
	\vspace{-3ex}
	\begin{center}
		\includegraphics[width =1\columnwidth]{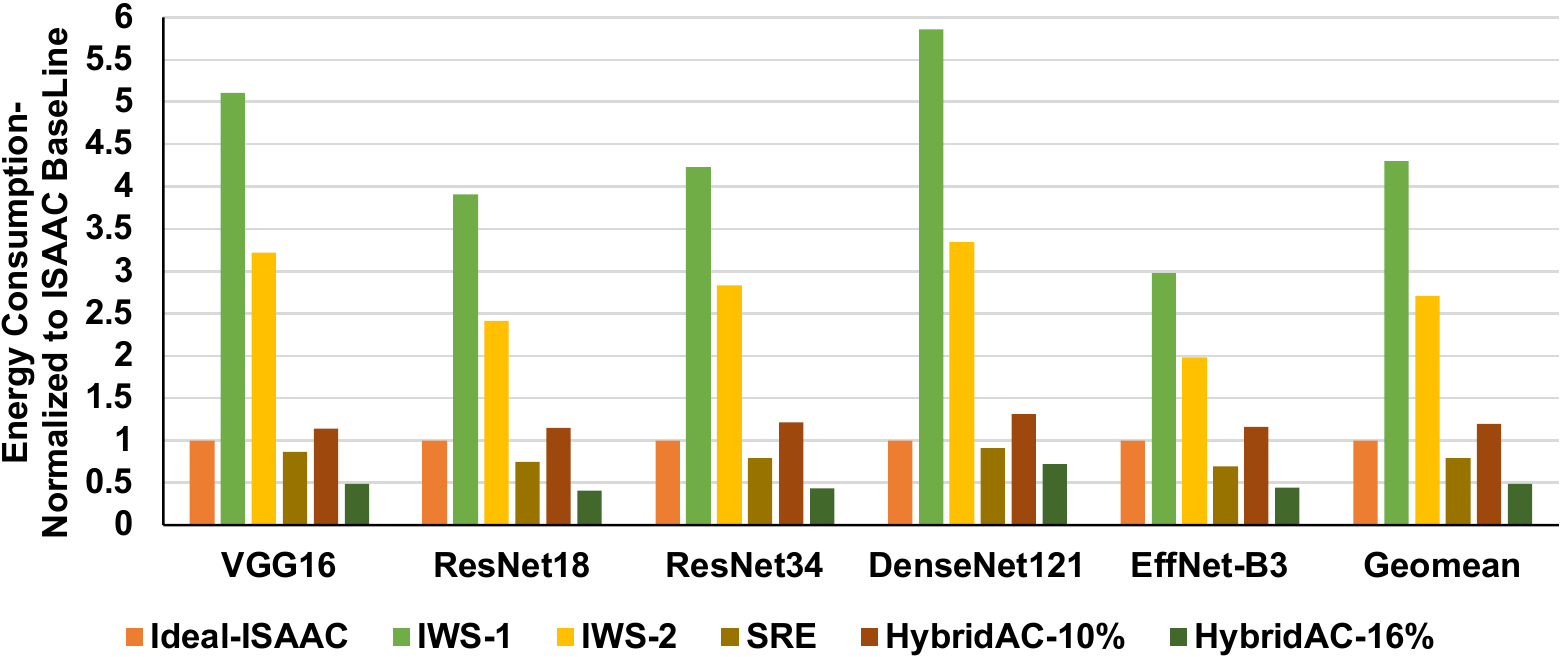}
	\vspace{-4ex}
	\caption{Energy consumption of different DNNs using different architectures
		\label{figure:energy}}
	\end{center}
	\vspace{-5ex}
\end{figure}

The results of execution time and energy of different DNNs using different methods over CIFAR100 dataset are demonstrated in Figures~\ref{figure:time} and~\ref{figure:energy} where idealistically assuming even ISAAC and SRE are immune to noise (ISO-accuracy). We mimic the behaviour and data flow of the evaluated architectures using our in-house simulator.  

On average, the execution time of Ideal-ISAAC is 3.6$\times$ and 1.6$\times$ better than IWS-1 and IWS-2~\cite{Dash2020,dash2021robust}, respectively. Due to exploiting activation and weight sparsity, SRE~\cite{RSE2019} can outperform ISAAC by 15$\times$; however, by having weight and activation as 8-bit rather than 16-bit, the sparsity exploitation is reduced. In addition, SRE has substantial column/row indexing overhead and relatively lower throughput as it activates only 16 rows. In HybridcAC-10\%, we assume up to 10\% of weights can be transferred to digital cores. Hence, if a design needs more percentage of weight to be mapped to the digital cores, it needs to wait until the existing digital cores become available for the next computation.

As the plots show, due to load unbalancing issue, the results of  HybridcAC-10\% are worse than ISAAC and SRE. Following our discussion in Section~\ref{sec:efficiency}, by moving more weights to the digital cores (up to 16\%), we can have a better load balancing that leads to improved timing and energy results. This amount best fits for the evaluated DNNs except DensNet121 over ImageNet as the results in Table~\ref{table:Prot_ChW_CIFAR} and Figure~\ref{figure:Prot_ChW_IamgeNet_ResNet18ResNet34DenseNet121} shows. That is why HybridcAC-16\% provides promising results. It improves the ISAAC timing by 26\% and SRE by 14\%. 
It is worth mentioning that having a perfect mapping and load-balancing for the mixed-signal designs is still an open challenge which is out of the scope of this work.

For power consumption, HybridAC consumes less power, has a shorter execution time. As we discussed in Section~\ref{sec:BB}, since the digital and analog cores do not share any input/ activations --while in the IWS baselines the same input data of analog needs to be replicated in digital units even if very few wights are transferred to digital cores--, HybridAC-16\% consumes less energy than the corresponding baselines. Since HybridAC-10\% has longer execution time, its energy consumption is more than HybridAC-16\% and SRE. 
In average, HybridAC-16\% improves energy consummation over Ideal-ISAAC, SRE, IWS-1, and IWS-2 by 52\%, 40\%, 8.9$\times$, and 5.6$\times$, respectively. Being faster and consuming less energy than the evaluated baselines leads to a better energy-delay product as well. It is noteworthy to mention that in practice, neither ISAAC, nor SRE are immune to weight variation and noise, which made them impractical for real-world application.

\subsubsection{Avoiding Accuracy Drop by Activating more Wordlines}

\begin{figure}[t!]
	\begin{center}
		\includegraphics[width =1\columnwidth]{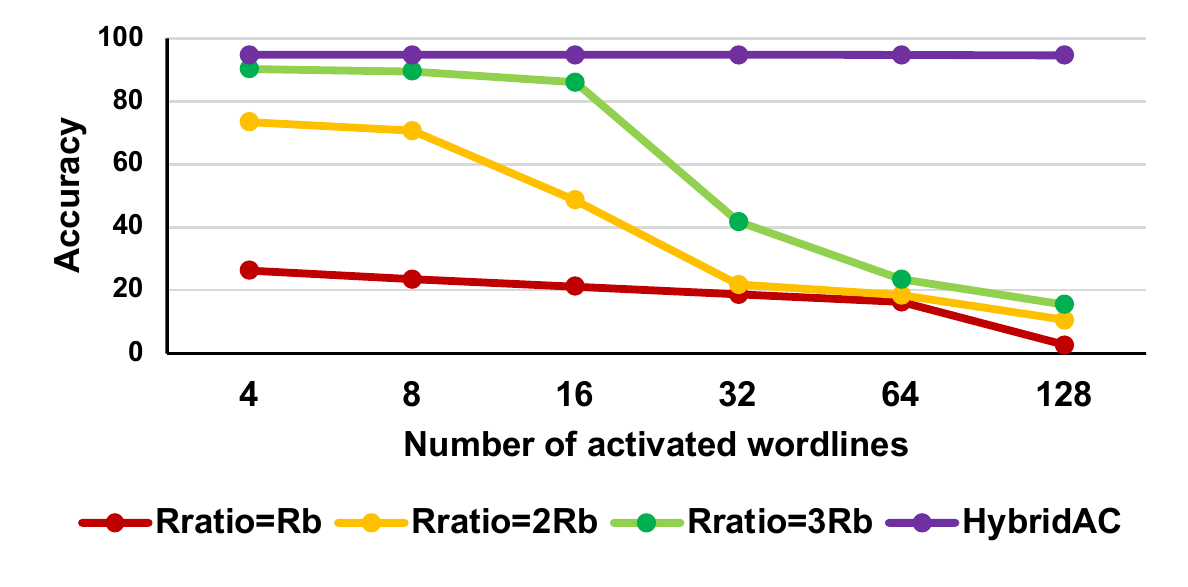}
	\vspace{-4ex}
	\caption{Accuracy degradation by increasing number of activated wordlines with no weight protection method 
	\label{figure:activatedWL2}}
	\end{center}
	\vspace{-5ex}
\end{figure}

As we discussed in Section~\ref{sec:intro}, existing mixed-signal designs especially those that rely on bias and offset subtraction method cannot activate many wordlines at the same time.
This problem can be mitigated if we can make the important weights immune against variations. Figure~\ref{figure:activatedWL2} shows our simulation study on different scenarios for ResNet18/ CIFAR10 where we increase the $R-{ratio}=\frac{R_{on}}{R_{off}}$ and decrease conductance deviation ($\sigma$). The first plot (e.g., red one) considers $R-{ratio}$ of VTEAM model~\cite{kvatinsky2015vteam} and $\sigma=50\%$ as a baseline ($R-{ratio}=R_b$). The second and third plots are for the situations when $R_{ratio}=2R_b$ $\sigma=\sigma_b/2$ and $R_{ratio}=3R_b$ $\sigma=\sigma_b/3$, respectively. The last one shows the behaviour of HybridAC.
These figures show that the amount of the accuracy degradation is considerable in practice while HybridAC can reduce such an accuracy drop to less than 0.9\%.

\subsubsection{Accuracy Improvement Studies}
Feinberg et al. propose an error correction scheme using AN-code to restore the accuracy loss~\cite{Feinberg2018}. AN codes rectify errors in linear functions and only consider the random telegraph noise. Moreover, it adds considerable hardware overheads, which makes it less scalable~\cite{jain2019cxdnn}.
Prior works activate fewer wordlines and rely on ADC quantization to surpass the propagated errors~\cite{zhang2020mitigate, yuan2021forms}. However, this solution cannot completely resolve the issue. Besides, their methods show lower throughput and area/power efficiency. 
The method proposed by Liu~\cite{liu2019fault} requires fine-tuning the network. This work replaces the final linear classification layer with a set of collaborative logistic classifiers to overcome the performance degradation, which needs retraining and add hardware overheads. In addition, modern image classifier architecture (e.g., SqueezeNet) does not incorporate any linear layer for making the final decision. 
Long et al. consider a Gaussian noise model and propose an algorithmic solution that needs retraining by incorporating the modeled noise~\cite{Long2019}.
Similarly, CxDNN utilizes a retraining method to recover accuracy loss due to the digital to analog conversion~\cite{jain2019cxdnn}. 

However, retraining-based approaches are less effective when models of variations assumed during retraining do not accurately match the variations experienced by the hardware during inference. In general, exact and comprehensive modeling of noise is a hard task that makes retraining-based methods less effective.

\section{Conclusion}
In HybridAC, for the first time, we propose a novel 
algorithm hardware framework that minimizes 
accuracy degradation, data movement, and energy 
consumption.
We propose an input channel-wise weight selection method, where the important channels along with their corresponding input data are moved to a novel tiny digital accelerator.
Our proposed method avoids input data replication and adding weight overheads. Besides, unlike the existing solutions, it does not require any retraining. In addition, it enables us to use a fewer number of crossbars, smaller peripheral circuitry, and ADCs with lower precision. The proposed solution also unlocks hybrid quantization where we have different quantization bits for different input channels per layer depending on where those channels are mapped to (i.e., analog or digital). 
We believe HybridAC gives an insight that having hybrid accelerator not only provides more immunity to noise and variations, but also promises a better area/power efficiency compared to the current mixed-signal accelerators.

\section*{ACKNOWLEDGEMENT}
This material is based on work supported in parts by National Science Foundation (\#1740197) and Semiconductor Research Corporation under the E2CDA program.

\bibliographystyle{IEEEtranS}
\bibliography{refs}

\end{document}